\documentclass[journal]{IEEEtran}

\usepackage{cite}
\usepackage{todonotes}
\usepackage{amsmath,amssymb,amsfonts}
\usepackage{algorithmic}
\usepackage{algorithm}
\usepackage{graphicx}
\usepackage{textcomp}
\usepackage{xcolor}
\usepackage{booktabs}
\usepackage{multirow}
\usepackage{array}
\usepackage[caption=false]{subfig}
\usepackage{tikz}
\usepackage{stfloats}
\usepackage{float}
\usepackage{xurl}
\usetikzlibrary{arrows.meta, positioning, fit, backgrounds}
\usetikzlibrary{shapes.geometric, calc, shadows}

\def\BibTeX{{\rm B\kern-.05em{\sc i\kern-.025em b}\kern-.08em
    T\kern-.1667em\lower.7ex\hbox{E}\kern-.125emX}}

\begin{document}

\title{Accelerating HEVC Intra Partitioning via a CNN-Hierarchical Attention Transformer Hybrid}

\author{Krishna~Kumar~Sharma
        and~Somdyuti~Paul%
\thanks{Krishna Kumar Sharma and Somdyuti Paul are with the Department of Artificial Intelligence,
Indian Institute of Technology Kharagpur, Kharagpur, West Bengal 721302, India
(e-mail: krishnakumarsharma0737@gmail.com; somdyuti@ai.iitkgp.ac.in)}%
}

%\markboth{IEEE Transactions on Circuits and Systems for Video Technology}%
%{Sharma \MakeLowercase{\textit{et al.}}: Accelerating HEVC Intra Encoding Using HFViT}

\maketitle

\begin{abstract}
The recursive quad-tree partitioning in High Efficiency Video Coding (HEVC) incurs considerable computational overhead, with exhaustive rate-distortion optimization for CTU partition prediction consuming the dominant share of encoding time. Although partition prediction through deep learning has emerged as a viable encoding accelerator, an architectural dichotomy remains largely unaddressed: CNNs are computationally efficient but spatially myopic due to their localized effective receptive fields, failing to capture long range semantic relationships and repetitive textures; conversely, transformer based architectures are better at capturing global context but incur prohibitive CPU latency, a critical liability that impedes deployment which is predominantly CPU-bound. This paper introduces Hybrid Fast Vision Transformer (HFViT), a hybrid architecture designed to accelerate HEVC intra-mode partition prediction. HFViT fuses a reparameterized depthwise-separable convolutional backbone with a Hierarchical Attention Transformer (HAT) mechanism, leveraging a carrier token scheme to enable efficient global information propagation at sub-quadratic complexity. Post-training structural fusion collapses batch normalization into preceding layers to further reduce latency. Comprehensive evaluation reveals the efficacy of HFViT in accelerating HEVC intra-encoding across resolutions. On standard JCT-VC test sequences, HFViT reduces the average VMAF BD-rate penalty by 2.4, 2.6, and 7.9 percentage points on Classes A, B and E, respectively, as compared to the competing ETH-CNN baseline while maintaining CPU inference latency within 8\% of the CNN baseline and surpassing it on GPU by 40\%, establishing practical viability for real-time encoder integration. 
\end{abstract}


\IEEEpeerreviewmaketitle

\section{Introduction}
\label{sec:introduction}

\IEEEPARstart{T}{he} explosive growth of high-resolution, high dynamic range and high frame rate
video content demands increasingly efficient compression technologies. High Efficiency Video Coding (HEVC/H.265) achieves approximately 50\% bitrate reduction compared to H.264/AVC while maintaining equivalent perceptual quality~\cite{sullivan2012overview}. However, this compression efficiency is achieved at the expense of significantly higher computational cost, with HEVC encoders exhibiting two to four times higher complexity than their H.264 predecessors~\cite{bossen2012hevc}. The primary computational bottleneck stems from HEVC's flexible Coding Unit (CU) partitioning mechanism. Unlike H.264's fixed macroblock structure, HEVC employs quad-tree partitioning where each 64$\times$64 Coding Tree Unit (CTU) is recursively subdivided into CUs of sizes \{64,32,16,8\}$\times$\{64,32,16,8\} across three levels \cite{sullivan2012overview}. The reference encoder performs rate-distortion optimization (RDO) \cite{rdo}, evaluating all feasible partition combinations which lie within a combinatorial search space. This brute-force search consumes 60--80\% of total encoding time~\cite{kim2012early,xiong2013fast}.

Traditional acceleration approaches for partitioning HEVC CUs employ heuristic-based early termination using handcrafted features such as texture variance, edge orientations,  motion vectors, etc. ~\cite{shen2014adaptive,zhang2015machine, grellert2018fast}. These methods achieve modest speedups in the range of 30--45\% at the expense of 1.5--3.0\% BD-rate penalties. Deep learning has transformed this landscape with convolutional neural networks (CNNs) automatically learning hierarchical representations, achieving over 50\% encoding time reduction~\cite{liu2016cu,xu2018reducing,li2017deep}. Moreover, CNN inference exhibits regular computational patterns enabling efficient GPU acceleration.

Despite these advances, CNNs face an inherent architectural limitation: their local receptive fields constrain global context modeling. Although the theoretical receptive field of CNNs progressively increases with the depth of the architecture, the pixels closer to the center contribute more to the output than those at the edges. This asymmetry causes the effective receptive field, which can be modeled by a Gaussian distribution, to be a fraction of the theoretical receptive field \cite{luo2016understanding}. Vision Transformer (ViT)~\cite{dosovitskiy2020image} addresses this through self-attention mechanisms that compute pairwise relationships across all tokens, extracted in the form of non-overlapping spatial patches. Irrespective of their spatial location within the image, all patches contribute equally to the output, thereby allowing transformer models to truly learn from the global context. While ViTs are highly parallelizable and very fast on GPUs due to the efficiency of parallelizing matrix multiplications required for attention computations that are conducted using highly optimized kernels, they remain significantly slower than CNNs on CPUs due to irregular memory access patterns, higher memory bandwidth requirements, and lack of equivalently optimized kernels. This is a crucial consideration for practical encoder integration, where inference is frequently CPU-bound.

This paper proposes \textbf{Hybrid Fast Vision Transformer (HFViT)}, a novel architecture that bridges the accuracy-efficiency gap in data-driven learning of HEVC CU partition prediction. Unlike CNNs that reason locally and standard vision transformers that reason expensively with quadratic complexity, HFViT employs hierarchical local-global attention mechanisms capturing both fine-grained texture details and long-range spatial dependencies at near-CNN computational cost. Through QP-conditioned feature modulation, HFViT makes partition decisions that adapt to both content complexity and target encoding quality, enabling rate-distortion (RD) aware predictions. Task-specific optimizations, including carrier token interactions for efficient global information exchange and batch normalization (BN) \cite{ioffe2015batch} layer fusion during deployment, allow HFViT to achieve superior RD performance compared to CNN baselines while maintaining competitive real-time encoding speed even when deployed without GPU acceleration.

Our key contributions can be summarized as follows:
\begin{itemize}
\item We conducted an extensive exploration of efficient transformer based architectures, systematically evaluating their computational complexity and inference latency on both CPU and GPU to identify architectures suitable for HEVC CU partition prediction.
\item We introduced a novel HFViT architecture specifically designed for HEVC intra-mode CU partition prediction, incorporating hierarchical attention with carrier token interaction and QP-aware feature modulation.
\item We integrated the trained HFViT model with the reference HEVC encoder to demonstrate HFViT's superior RD performance over CNN baselines across different quality metrics with competitive encoding speed across multiple resolutions. 
\footnote{To facilitate reproducibility, the source code and trained models are made publicly available at \url{https://github.com/Krishna737Sharma/HFViT_CU_Partitioning_in_HEVC}.}
\end{itemize}

The rest of the paper is organized as follows. Section II reviews relevant prior works. Section III describes the proposed HFViT architecture. Section IV presents the experimental results. Finally, Section V concludes the paper and provides directions for extending the work.

\section{Related Work}
\label{sec:related}

Building upon the acceleration approaches for partitioning CUs outlined earlier, we review specific methodological advances and identify the critical gaps motivating our work. Among heuristic methods, inter-level correlations for adaptive mode decisions were exploited in \cite{shen2014adaptive}, while texture homogeneity with gradient analysis was used in \cite{zhang2015machine}. In \cite{kim2012early} an RD cost-based early termination was proposed, and in \cite{xiong2013fast} motion divergence metrics were introduced for inter-CU selection. These methods rely on manually designed thresholds requiring content-specific tuning.

Among learned approaches, \cite{liu2016cu} pioneered learned CU partition decision using shallow CNNs, which was subsequently extended with deeper architectures and asymmetric convolutions in \cite{xu2018reducing}, forming the ETH-CNN baseline used throughout our evaluation. Multi-scale feature fusion matching the CU hierarchy was used in \cite{li2017deep}. Lightweight CNN-based learned partition prediction has also been introduced for partitioning VP9 superblocks using a bottom-up block merge prediction mechanism ~\cite{paul2020speeding} and using shape adaptive pooling layers to partition CUs in VVC~\cite{tang2019adaptive}.

With the introduction of Vision Transformer (ViT)~\cite{dosovitskiy2020image}, transformer based neural network architectures were shown to achieve superior performance as compared to CNN based architectures on several vision tasks. Subsequent efforts to reduce the quadratic complexity of attention computation have produced notable efficient architectural innovations. Swin Transformer~\cite{liu2021swin} introduced shifted window attention achieving linear complexity. LeViT~\cite{graham2021levit} combined convolutional stems with attention-based downsampling. MobileFormer~\cite{chen2022mobile} proposed parallel CNN-transformer branches with bidirectional bridging. EfficientViT~\cite{liu2023efficientvit} employed cascaded group attention for memory efficiency. FastViT~\cite{vasu2023fastvit} utilized structural reparameterization for accelerated inference. FasterViT~\cite{hatamizadeh2023fastervit} introduced carrier tokens enabling hierarchical global context propagation. EfficientFormer~\cite{li2022efficientformer} achieved inference speeds comparable to MobileNet \cite{howard2017mobilenets} through a latency-driven design paradigm. Recent work has also explored applying transformer-based models to HEVC's successor standard Versatile Video Coding (VVC/H.266)~\cite{bross2021overview}, where more complex multi-type tree partitioning further increases the benefit of global context modeling~\cite{li2025deep}. However, none of these architectures address video coding-specific requirements or CPU-bound latency constraints to achieve inference speeds that are competitive with CNN baselines.

To quantify this efficiency gap and motivate the need for HFViT, we systematically adapted several popular transformer architectures designed to reduce complexity of attention computation for the purpose of CU partition prediction, and benchmarked their inference latency on both CPU and GPU. Table~\ref{tab:model_comparison} presents the inference times of each architecture. The inference time of the ETH-CNN model is also shown for reference. The same batch size of 64 was used for evaluating the latency of all models and all the architectures were designed to have similar learning capacity in terms of the number of trainable parameters to facilitate a fair comparison.

\begin{table}[htbp]
\centering
\caption{Comparative Analysis of Transformer Architectures for HEVC CTU Partition Prediction (BS\,64)}
\label{tab:model_comparison}
{\fontsize{7.0}{8.5}\selectfont
\setlength{\tabcolsep}{3.5pt}
\renewcommand{\arraystretch}{1.1}
\begin{tabular}{l r c c r}
\toprule
\textbf{Model} & \textbf{Params} &
\textbf{\shortstack{CPU Latency \\ (Time\,/\,CTU)}} &
\textbf{\shortstack{GPU Latency \\ (Time\,/\,CTU)}} &
\textbf{GFLOPS} \\
\midrule
ViT~\cite{dosovitskiy2020image}             & 1,279,317 & \,4.77\,ms   & \,0.109\,ms  & 0.176 \\
Swin-T~\cite{liu2021swin}          & 1,087,578 & \,1.73\,ms   & \,0.188\,ms  & 0.034 \\
LeViT~\cite{graham2021levit}             & 1,081,461 & \,1.11\,ms   & \,0.109\,ms  & 0.030 \\
MobileFormer~\cite{chen2022mobile}     & 1,172,597 & \,1.09\,ms   & \,0.344\,ms  & 0.009 \\
EfficientFormer~\cite{li2022efficientformer} &   969,861 & \,3.02\,ms   & \,0.125\,ms  & 0.029 \\
EfficientViT~\cite{liu2023efficientvit}    & 1,430,770 & \,0.50\,ms   & \,0.109\,ms  & 0.009 \\
FastViT~\cite{vasu2023fastvit}          &   957,155 & \,6.41\,ms   & \,0.313\,ms  & 0.206 \\
FasterViT~\cite{hatamizadeh2023fastervit}        & 1,373,725 & \,0.59\,ms   & \,0.062\,ms  & 0.014 \\
\midrule
ETH-CNN~\cite{xu2018reducing}
                & 1,288,210 &  \,0.25\,ms   & \,0.078\,ms  & 0.003 \\
\textbf{HFViT (Ours)}
                & \textbf{1,670,794} & \textbf{\,0.27\,ms} & \textbf{\,0.047\,ms} & \textbf{0.0048} \\
\bottomrule
\end{tabular}
}
\end{table}

While comparing models with similar capacity, we note that transformer based architectures have significantly higher computational complexity as compared to the ETH-CNN model as reported in units of Giga Floating-Point Operations Per Second (GFLOPS) in the last column of Table \ref{tab:model_comparison}. The consequent CPU latency is the critical bottleneck for practical encoder integration, as HEVC encoders are predominantly deployed in CPU-bound pipelines. On CPU, transformers exhibit substantially higher latency than CNNs due to limitations of caching mechanisms, higher memory bandwidth requirements, and the absence of heavily optimized kernels comparable to those available for convolutions. Even the fastest competing transformer, EfficientViT, incurs $0.50$\,ms/CTU, which is twice the ETH-CNN baseline of $0.25$\,ms/CTU. ViT ($4.77$\,ms) and Swin-T ($1.73$\,ms) are slower still by margins that would negate any encoding time savings. This persistent CPU efficiency gap is the primary obstacle preventing transformer adoption in real-time encoding pipelines.

GPU latency presents a more favorable picture for transformers, owing to heavily optimized attention kernels and the inherently parallel nature of self-attention computation. On GPU, most efficient transformers (ViT, LeViT, EfficientViT) match or approach the ETH-CNN baseline of $0.078$\,ms/CTU at approximately $0.109$\,ms/CTU. However, GPU-bound inference is less representative of practical deployment scenarios where inference is frequently CPU-bound.

Table \ref{tab:model_comparison} shows that FasterViT provides the most promising trade-off between CPU and GPU latencies, which motivated the design of the proposed HFViT model. Building upon FasterViT~\cite{hatamizadeh2023fastervit} and FastViT~\cite{vasu2023fastvit}, HFViT bridges the CPU efficiency gap through targeted, domain-specific architectural adaptations. Instead of employing deep designs tailored for high-resolution RGB imagery, HFViT adopts a streamlined hierarchical structure with channel dimensions specifically optimized for low-resolution, single-channel CTU inputs. Furthermore, HFViT applies post-training BN \cite{ioffe2015batch} fusion, wherein BN parameters are absorbed into the preceding layers to eliminate runtime normalization overhead. This structural reparameterization reduces inference-time computational overhead without affecting model performance. As a result, HFViT achieves a CPU latency of $0.27$\,ms/CTU, which is within $8\%$ of the ETH-CNN baseline ($0.25$\,ms/CTU), while attaining a GPU latency of $0.047$\,ms/CTU, outperforming ETH-CNN ($0.078$\,ms/CTU) by approximately $40\%$. Consequently, HFViT demonstrates practical viability for real-time CTU partition prediction in both CPU and GPU-based encoding pipelines, representing a significant advancement over prior CNN-based models that lack global context modeling and earlier transformer architectures that incur prohibitive CPU latencies.

\section{Proposed HFViT Architecture}
\label{sec:method}

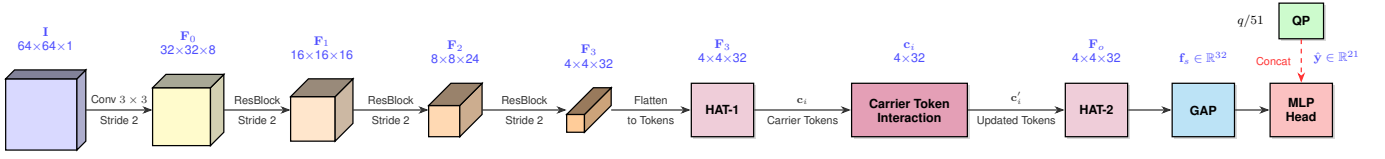
\begin{figure*}[!t]
\centering
\resizebox{\textwidth}{!}{
\begin{tikzpicture}[
    node distance=0.8cm and 1.2cm,
    module/.style={
        rectangle, draw=black, thick,
        minimum height=1.2cm, minimum width=1.4cm,
        align=center, font=\footnotesize\sffamily\bfseries,
        fill opacity=0.9, text opacity=1,
        drop shadow={shadow xshift=0.5mm, shadow yshift=-0.5mm, opacity=0.3}
    },
    arrow/.style={
        ->, >=Stealth, thick, draw=black!80
    },
    ar_label/.style={
        font=\scriptsize\sffamily, text=black!80, align=center
    },
    ar_label_below/.style={
        font=\scriptsize\sffamily, text=black!80, align=center
    },
    block_name/.style={
        font=\footnotesize\sffamily\bfseries, align=center
    },
    block_dim/.style={
        font=\footnotesize\sffamily, align=center, text=blue!70
    }
]

% ---------- FIXED CUBOID ----------
\tikzset{
  pics/cuboid/.style args={w=#1, h=#2, d=#3, fill=#4}{
    code={
      \colorlet{basecolor}{#4}
      \colorlet{rightcolor}{basecolor!80!black}
      \colorlet{topcolor}{basecolor!60!black}

      \begin{scope}[shift={(-#1/2 - #3/2, -#2/2 - #3/2)}]
        \draw[fill=basecolor, thick] (0,0) -- ++(#1,0) -- ++(0,#2) -- ++(-#1,0) -- cycle;
        \draw[fill=rightcolor, thick] (#1,0) -- ++(#3,#3) -- ++(0,#2) -- ++(-#3,-#3) -- cycle;
        \draw[fill=topcolor, thick] (0,#2) -- ++(#3,#3) -- ++(#1,0) -- ++(-#3,-#3) -- cycle;

        \coordinate (-west)   at (0, #2/2 + #3/2);
        \coordinate (-east)   at (#1+#3, #2/2 + #3/2);
        \coordinate (-top)    at (#1/2+#3/2, #2+#3);
        \coordinate (-bottom) at (#1/2+#3/2, 0);
        \coordinate (-center) at (#1/2+#3/2, #2/2+#3/2);
      \end{scope}
    }
  }
}

% ---------- BLOCKS ----------
\pic (ctu)  at (0,0)    {cuboid={w=1.6, h=1.6, d=0.2, fill=blue!15}};
\pic (stem) at (3.2,0)  {cuboid={w=1.3, h=1.3, d=0.3, fill=yellow!25}};
\pic (stg1) at (6.2,0)  {cuboid={w=1.0, h=1.0, d=0.4, fill=orange!20}};
\pic (stg2) at (9.2,0)  {cuboid={w=0.7, h=0.7, d=0.5, fill=orange!30}};
\pic (stg3) at (12.2,0) {cuboid={w=0.4, h=0.4, d=0.6, fill=orange!40}};

\node[module, fill=purple!20] (hat1)    at (15.2, 0) {HAT-1};
\node[module, fill=purple!40, minimum width=2.6cm] (ctinter) at (19.4, 0) {Carrier Token\\Interaction};
\node[module, fill=purple!20] (hat2)    at (23.6, 0) {HAT-2};
\node[module, fill=cyan!25]   (gap)     at (26.0, 0) {GAP};
\node[module, fill=red!25]    (head)    at (28.2, 0) {MLP\\Head};

\node[module, fill=green!20, minimum height=0.8cm, minimum width=1.0cm] (qp) at (28.2, 2.0) {QP};

% ---------- DIMENSIONS ----------
\node[block_dim, above=0.3cm of ctu-top]      {$\mathbf{I}$\\64$\times$64$\times$1};
\node[block_dim, above=0.3cm of stem-top]     {$\mathbf{F}_0$\\32$\times$32$\times$8};
\node[block_dim, above=0.3cm of stg1-top]     {$\mathbf{F}_1$\\16$\times$16$\times$16};
\node[block_dim, above=0.3cm of stg2-top]     {$\mathbf{F}_2$\\8$\times$8$\times$24};
\node[block_dim, above=0.3cm of stg3-top]     {$\mathbf{F}_3$\\4$\times$4$\times$32};

\node[block_dim, above=0.3cm of hat1.north]   {$\mathbf{F}_3$\\4$\times$4$\times$32};
\node[block_dim, above=0.3cm of ctinter.north]{$\mathbf{c}_i$\\4$\times$32};
\node[block_dim, above=0.3cm of hat2.north]   {$\mathbf{F}_o$\\4$\times$4$\times$32};
\node[block_dim, above=0.3cm of gap.north]    {$\mathbf{f}_s \in \mathbb{R}^{32}$};
\node[block_dim, above=0.3cm of head.north, xshift=0.8cm] {$\hat{\mathbf{y}} \in \mathbb{R}^{21}$};

\node[block_name, left=0.2cm of qp.west] {$q/51$};

% ---------- ARROWS (FIXED) ----------
\draw[arrow] (ctu-east) -- 
node[ar_label, above] {Conv $3\times3$}
node[ar_label_below, below] {Stride 2}
(stem-west);

\draw[arrow] (stem-east) -- 
node[ar_label, above] {ResBlock}
node[ar_label_below, below] {Stride 2}
(stg1-west);

\draw[arrow] (stg1-east) -- 
node[ar_label, above] {ResBlock}
node[ar_label_below, below] {Stride 2}
(stg2-west);

\draw[arrow] (stg2-east) -- 
node[ar_label, above] {ResBlock}
node[ar_label_below, below] {Stride 2}
(stg3-west);

\draw[arrow] (stg3-east) -- 
node[ar_label, above] {Flatten}
node[ar_label_below, below] {to Tokens}
(hat1.west);

\draw[arrow] (hat1.east) -- 
node[ar_label, above] {$\mathbf{c}_i$}
node[ar_label_below, below] {Carrier Tokens}
(ctinter.west);

\draw[arrow] (ctinter.east) -- 
node[ar_label, above] {$\mathbf{c}_i'$}
node[ar_label_below, below] {Updated Tokens}
(hat2.west);

\draw[arrow] (hat2.east) -- (gap.west);
\draw[arrow] (gap.east) -- (head.west);

% QP connection
\draw[arrow, dashed, draw=red!80] 
(qp.south) -- 
node[midway, left=3pt, font=\scriptsize\sffamily, text=red!80] {Concat}
(head.north);

\end{tikzpicture}
}
\caption{HFViT architecture: hierarchical convolutional stages extract multi-scale features $\mathbf{F}_0$--$\mathbf{F}_3$, followed by dual HAT blocks with carrier token interaction for efficient global context modeling. QP is injected at the classification head for rate-distortion-aware prediction.}
\label{fig:architecture}
\end{figure*}

\subsection{Overview and Design Rationale}

The overall HFViT pipeline, illustrated in Fig. ~\ref{fig:architecture}, consists of (1) a convolutional embedding stem incorporating four progressive downsampling stages, followed by (2) dual Hierarchical Attention Transformer (HAT) blocks with carrier token interaction, and a (3) QP-conditioned classification head. HFViT is primarily motivated by the FasterViT architecture~\cite{hatamizadeh2023fastervit}, which introduced the HAT mechanism with carrier tokens for efficient global context propagation. We combined the HAT mechanism with several task-oriented adaptations that were introduced for low-latency CU partition prediction as follows:

\begin{itemize}

\item \textbf{Depthwise Separable Lightweight Convolutional Stem:} while FasterViT features a convolutional stem for feature extraction, we designed the convolutional stem using depthwise separable convolutions \cite{chen2022mobile} to reduce the number of trainable parameters as well as computational complexity. Further, as the input is a low-resolution, single-channel luminance block, we employed a conservative channel progression of $8 \rightarrow 16 \rightarrow 24 \rightarrow 32$, unlike convolutional feature extraction stages designed for vision tasks on higher resolution RGB images that adopt more aggressive channel expansions. This tailored allocation avoids the severe overfitting that occurs when applying such models directly to simple inputs, substantially reducing the parameter count while preserving representational capacity.

\item \textbf{Shallower Attention Hierarchy:} since HEVC CTUs have a fixed, relatively small spatial resolution of $64\times64$, a deep HAT architecture is unnecessary. Thus, we created a shallower attention hierarchy comprising two HAT layers that capture global context while drastically reducing computational overhead.

\item \textbf{QP-Conditioned Prediction Head:} as partition decisions depend on the quantization parameter (QP), we introduced explicit QP conditioning directly into the classification head. This allows a single model to make dynamic, rate-distortion-aware partition decisions that adaptively scale with the target encoding quality.

\item \textbf{Batch Norm Fusion:} inspired by FastViT~\cite{vasu2023fastvit}, our architecture utilizes BN fusion during inference in two places: (i) in the depthwise separable convolutional stem, and (ii) in the fully connected layers of the classification head. During training, the convolution-BN and linear-BN structures are decoupled, and BN provides extra trainable variables that stabilize gradients and improve optimization. During inference, structural reparameterization is applied to mathematically fuse the BN parameters directly into the weights of the preceding convolutional or fully connected layers. This yields a single, highly efficient operation in each case, eliminating BN latency entirely during deployment.

\end{itemize}

The architecture is termed \textit{hybrid} because it strategically merges two distinct design philosophies. It borrows the CNN-driven computational efficiency and structural reparameterization, and fuses it with the global reasoning capability of hierarchical attention mechanism facilitated by carrier tokens. This tailored combination bridges the critical gap between the ultra-low CPU latency required for practical HEVC encoding and the robust spatial context modeling necessary for accurate CU partition prediction.

\subsection{Problem Formulation}

Given a CTU luminance (Y-channel) block $\mathbf{I} \in \{0, 1, \ldots, 255\}^{64 \times 64}$ and quantization parameter $q \in \{0, 1, \cdots, 51\}$, HFViT predicts optimal quad-tree partitions across three depth levels without exhaustive RDO. The input $\mathbf{I}$ is normalized to $[0, 1]$ before processing. At depth $d \in \{0, 1, 2\}$ for CU sizes $\{64\times64, 32\times32, 16\times16\}$, HFViT predictions are:
\begin{itemize}
\item \textbf{Depth 0:} One decision $\mathbf{y}_0 \in \{0,1\}$
\item \textbf{Depth 1:} Four decisions $\mathbf{y}_1 \in \{0,1\}^4$
\item \textbf{Depth 2:} Sixteen decisions $\mathbf{y}_2 \in \{0,1\}^{16}$
\end{itemize}

The complete target is $\mathbf{y} = [\mathbf{y}_0, \mathbf{y}_1, \mathbf{y}_2]^T \in \{0,1\}^{21}$. Our model learns:
\begin{equation}
f_\theta: (\mathbf{I}, q) \rightarrow \hat{\mathbf{y}} \in [0,1]^{21}
\end{equation}

The stagewise components of HFViT are elucidated in the subsequent paragraphs.
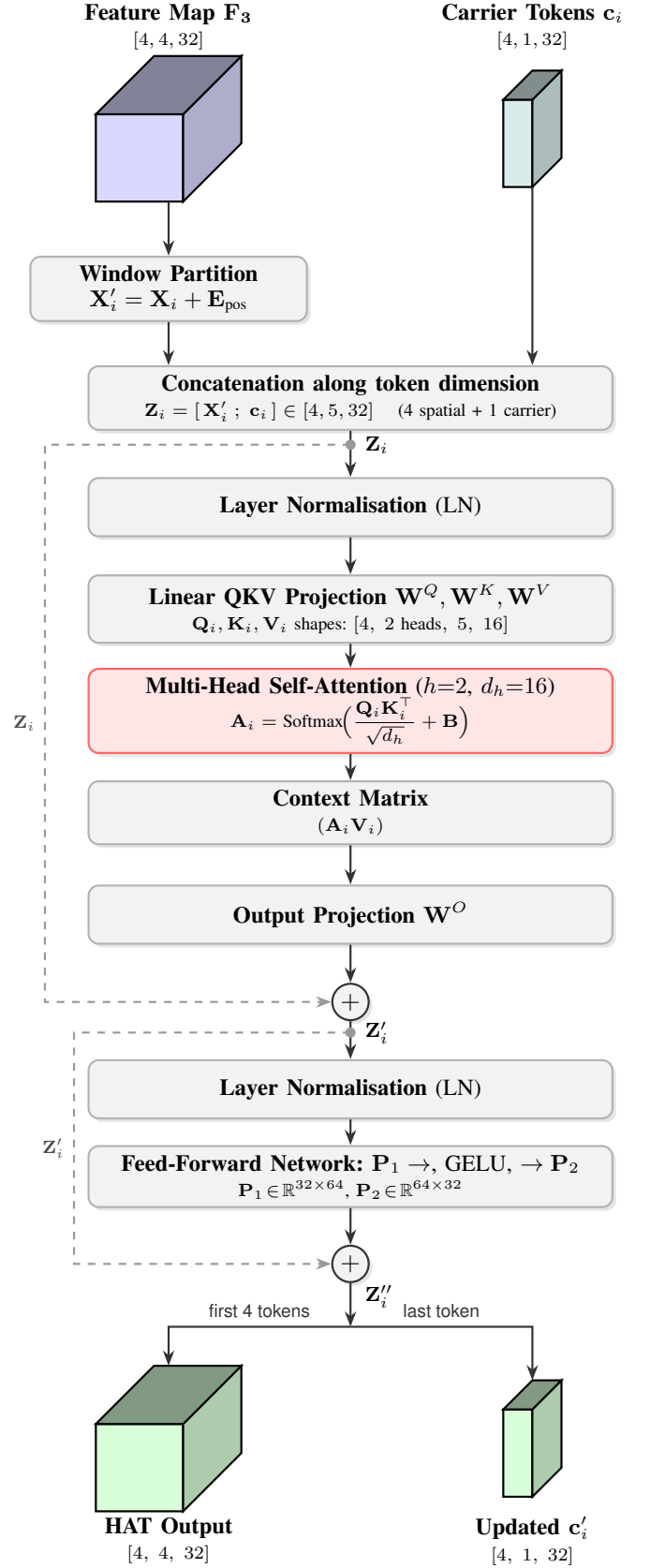
\begin{figure}[htbp]
\centering
\resizebox{\columnwidth}{!}{%
\begin{tikzpicture}[
  node distance=0.6cm and 0.5cm,
  process/.style={rectangle, draw=gray!60, fill=gray!10, thick, rounded corners=4pt, minimum height=0.8cm, align=center, font=\small, drop shadow={shadow xshift=0.5mm, shadow yshift=-0.5mm, opacity=0.2}, minimum width=7.2cm},
  attn/.style={rectangle, draw=red!60, fill=red!10, thick, rounded corners=4pt, minimum height=0.8cm, align=center, font=\small, drop shadow={shadow xshift=0.5mm, shadow yshift=-0.5mm, opacity=0.2}, minimum width=7.2cm},
  arrow/.style={-{Stealth[length=6pt]}, thick, draw=black!80},
  skip_arrow/.style={-{Stealth[length=6pt]}, thick, dashed, draw=gray!80},
  lbl/.style={font=\scriptsize\sffamily, text=black!70, align=center},
  add_node/.style={circle, draw=black!80, thick, fill=gray!10, minimum size=0.5cm, inner sep=0pt, font=\normalsize\bfseries, drop shadow={shadow xshift=0.5mm, shadow yshift=-0.5mm, opacity=0.2}}
]
\tikzset{
  pics/cuboid/.style args={w=#1, h=#2, d=#3, fill=#4}{
    code={
      % --- FIX: avoid chained color mixing ---
      \colorlet{basecolor}{#4}
      \colorlet{rightcolor}{basecolor!80!black}
      \colorlet{topcolor}{basecolor!60!black}

      \begin{scope}[shift={(-#1/2 - #3/2, -#2/2 - #3/2)}]
        \draw[fill=basecolor, thick] 
          (0,0) -- ++(#1,0) -- ++(0,#2) -- ++(-#1,0) -- cycle;

        \draw[fill=rightcolor, thick] 
          (#1,0) -- ++(#3,#3) -- ++(0,#2) -- ++(-#3,-#3) -- cycle;

        \draw[fill=topcolor, thick] 
          (0,#2) -- ++(#3,#3) -- ++(#1,0) -- ++(-#3,-#3) -- cycle;

        \coordinate (-west)   at (0, #2/2 + #3/2);
        \coordinate (-east)   at (#1+#3, #2/2 + #3/2);
        \coordinate (-top)    at (#1/2+#3/2, #2+#3);
        \coordinate (-bottom) at (#1/2+#3/2, 0);
        \coordinate (-center) at (#1/2+#3/2, #2/2+#3/2);
      \end{scope}
    }
  }
}
\node[font=\normalsize\bfseries] (title) at (0, 2.5) {HAT Block Internal Process (HAT-1 / HAT-2)};
\pic (fm) at (-2.5, 0) {cuboid={w=1.2, h=1.2, d=0.8, fill=blue!15}};
\node[align=center, font=\small] (fm_lbl) at (-2.5, 1.4) {\textbf{Feature Map} $\mathbf{F_3}$\\{\scriptsize $[4, 4, 32]$}};
\pic (ct_in) at (2.5, 0) {cuboid={w=0.4, h=1.2, d=0.4, fill=teal!15}};
\node[align=center, font=\small] (ct_lbl) at (2.5, 1.4) {\textbf{Carrier Tokens} $\mathbf{c}_i$\\{\scriptsize $[4, 1, 32]$}};

\node[process, minimum width=3.8cm] (winpart) at (-2.5, -2.2) {\textbf{Window Partition}\\ $\mathbf{X}_i' = \mathbf{X}_i + \mathbf{E}_{\text{pos}}$};

\node[process] (concat) at (0, -3.7) {\textbf{Concatenation along token dimension}\\{\scriptsize $\mathbf{Z}_i = [\,\mathbf{X}_i'\;;\;\mathbf{c}_i\,] \in [4, 5, 32]$ \quad (4 spatial + 1 carrier)}};

\node[process] (ln1) at (0, -5.2) {\textbf{Layer Normalisation} (LN)};
\node[process] (qkv) at (0, -6.6) {\textbf{Linear QKV Projection} $\mathbf{W}^Q, \mathbf{W}^K, \mathbf{W}^V$\\{\scriptsize $\mathbf{Q}_i,\mathbf{K}_i,\mathbf{V}_i$ shapes: $[4,\; 2\text{ heads},\; 5,\; 16]$}};
\node[attn] (attn) at (0, -8.0) {\textbf{Multi-Head Self-Attention} ($h{=}2$, $d_h{=}16$)\\{\scriptsize $\mathbf{A}_i = \text{Softmax}\!\Big(\dfrac{\mathbf{Q}_i\mathbf{K}_i^\top}{\sqrt{d_h}} + \mathbf{B}\Big)$}};
\node[process] (context) at (0, -9.4) {\textbf{Context Matrix}\\{\scriptsize $(\mathbf{A}_i\mathbf{V}_i)$}};
\node[process] (proj) at (0, -10.8) {\textbf{Output Projection} $\mathbf{W}^O$};

\node[add_node] (add1) at (0, -12.0) {$+$};

\node[process] (ln2) at (0, -13.2) {\textbf{Layer Normalisation} (LN)};
\node[process] (ffn) at (0, -14.4) {\textbf{Feed-Forward Network:} $\mathbf{P}_1 \rightarrow$, GELU, $\rightarrow \mathbf{P}_2$\\{\scriptsize $\mathbf{P}_1\!\in\!\mathbb{R}^{32\times64}$, $\mathbf{P}_2\!\in\!\mathbb{R}^{64\times32}$}};

\node[add_node] (add2) at (0, -15.6) {$+$};

\pic (out_x) at (-2.5, -18.0) {cuboid={w=1.2, h=1.2, d=0.8, fill=green!15}};
\node[align=center, font=\small] (out_x_lbl) at (-2.5, -19.4) {\textbf{HAT Output} \\{\scriptsize $[4,\;4, \;32]$}};
\pic (out_ct) at (2.5, -18.0) {cuboid={w=0.4, h=1.2, d=0.4, fill=green!15}};
\node[align=center, font=\small] (out_ct_lbl) at (2.5, -19.4) {\textbf{Updated $\mathbf{c}_i'$}\\{\scriptsize $[4,\;1,\;32]$}};

\draw[arrow] (fm-bottom) -- (winpart.north);
\draw[arrow] (winpart.south) -- ++(0,-0.4) -| (concat.north -| winpart.south);
\draw[arrow] (ct_in-bottom) -- (ct_in-bottom |- winpart.south) -- ++(0,-0.4) -| (concat.north -| ct_in-bottom);

\draw[arrow] (concat.south) -- coordinate[pos=0.3] (res1_start) node[right=2pt, pos=0.3, font=\footnotesize\sffamily] {$\mathbf{Z}_i$} (ln1.north);
\draw[arrow] (ln1.south) -- (qkv.north);
\draw[arrow] (qkv.south) -- (attn.north);
\draw[arrow] (attn.south) -- (context.north);
\draw[arrow] (context.south) -- (proj.north);
\draw[arrow] (proj.south) -- (add1.north);

\draw[arrow] (add1.south) -- coordinate[pos=0.3] (res2_start) node[right=2pt, pos=0.3, font=\footnotesize\sffamily] {$\mathbf{Z}_i'$} (ln2.north);
\draw[arrow] (ln2.south) -- (ffn.north);
\draw[arrow] (ffn.south) -- (add2.north);

\draw[arrow] (add2.south) -- node[right=2pt, pos=0.3, font=\footnotesize\sffamily] {$\mathbf{Z}_i''$} ++(0,-0.6) coordinate (split_pt);

\draw[arrow] (split_pt) -| (out_x-top) node[pos=0.25, above, font=\scriptsize\sffamily, text=black!80] {first 4 tokens};
\draw[arrow] (split_pt) -| (out_ct-top) node[pos=0.25, above, font=\scriptsize\sffamily, text=black!80] {last token};

\fill[gray!80] (res1_start) circle (2pt);
\draw[skip_arrow] (res1_start) -- ++(-4.2,0) |- node[lbl, left, pos=0.25] {$\mathbf{Z}_i$} (add1.west);

\fill[gray!80] (res2_start) circle (2pt);
\draw[skip_arrow] (res2_start) -- ++(-3.8,0) |- node[lbl, left, pos=0.25] {$\mathbf{Z}_i'$} (add2.west);

\end{tikzpicture}%
}
\caption{Internal structure of a HAT block. Spatial tokens and a carrier token are concatenated, processed through multi-head self-attention, a context matrix, and a feed-forward network with residual connections, then split back into updated spatial features and carrier token.}
\label{fig:hat_block_detail}
\end{figure}
\subsection{Depthwise Separable Convolution}

The convolutional stem of HFViT consists of four depthwise-separable convolution \cite{chollet2017xception} stages, progressively downsampling the input from $64{\times}64$ to $4{\times}4$ while expanding the channel dimensions following $[n_0, n_1, n_2, n_3] = [8, 16, 24, 32]$.

For the first stage ($\ell{=}0$), there is no residual block; the stride-2 depthwise kernel $\mathbf{U}^{d}_0 \in \mathbb{R}^{1 \times 3 \times 3}$ acts directly on the normalised single-channel input image $\mathbf{I}$, followed by the pointwise projection $\mathbf{U}^{p}_0 \in \mathbb{R}^{n_0 \times 1 \times 1 \times 1}$: 
\begin{align}
\mathbf{F}_0 &= \text{GELU}\!\left(\text{BN}\!\left(\mathbf{U}^{p}_0 * (\mathbf{U}^{d}_0 * \mathbf{I})\right)\right)
\end{align}

For stages $\ell \in \{1, 2, 3\}$, a residual block first processes $\mathbf{F}_{\ell-1}$ while preserving its channel dimension $n_{\ell-1}$. A depthwise kernel $\mathbf{W}^{d}_\ell \in \mathbb{R}^{n_{\ell-1} \times 3 \times 3}$ is applied independently per channel, followed by a pointwise kernel $\mathbf{W}^{p}_\ell \in \mathbb{R}^{n_{\ell-1} \times n_{\ell-1} \times 1 \times 1}$, with BN \cite{ioffe2015batch} and GELU activation \cite{hendrycks2016gaussian} (where $\text{GELU}(x) = x\Phi(x)$, and $\Phi(x)$ is the cumulative distribution function of the standard normal distribution):
\begin{align}
\mathbf{F}_\ell' &= \mathbf{F}_{\ell-1} + \text{GELU}\!\left(\text{BN}\!\left(\mathbf{W}^{p}_\ell * (\mathbf{W}^{d}_\ell * \mathbf{F}_{\ell-1})\right)\right)
\end{align}

A stride-2 depthwise convolution $\mathbf{U}^{d}_\ell \in \mathbb{R}^{n_{\ell-1} \times 3 \times 3}$ then halves the spatial resolution, while a pointwise projection $\mathbf{U}^{p}_\ell \in \mathbb{R}^{n_\ell \times n_{\ell-1} \times 1 \times 1}$ expands the channel dimension from $n_{\ell-1}$ to $n_\ell$, producing $\mathbf{F}_\ell$:
\begin{align}
\mathbf{F}_\ell &= \text{GELU}\!\left(\text{BN}\!\left(\mathbf{U}^{p}_\ell * (\mathbf{U}^{d}_\ell * \mathbf{F}_\ell')\right)\right)
\end{align}

Consequently, the sizes of the feature maps at each successive downsampling step are $\mathbf{F}_0 \in \mathbb{R}^{32\times32\times8}$, $\mathbf{F}_1 \in \mathbb{R}^{16\times16\times16}$, $\mathbf{F}_2 \in \mathbb{R}^{8\times8\times24}$, and $\mathbf{F}_3 \in \mathbb{R}^{4\times4\times32}$, respectively.

\subsection{Hierarchical Attention Transformer}

Following FasterViT~\cite{hatamizadeh2023fastervit}, we employed the HAT mechanism with carrier tokens to enable efficient global context exchange without the quadratic complexity of full self-attention. Here, the term \textit{hierarchical} refers to concurrently modeling spatial relationships at distinct semantic levels: fine-grained local spatial relationships (between image patches within a window) and local-to-global information exchange (via carrier tokens that facilitate global communication in the subsequent interaction layer).

\subsubsection{Carrier Token Mechanism}

We partition $\mathbf{F}_3 \in \mathbb{R}^{4\times4\times32}$ into $4$ non-overlapping windows of size $2\times2$. Each window generates a carrier token via adaptive average pooling:
\begin{equation}
\mathbf{c}_i = \text{AdaptiveAvgPool}(\mathbf{F}_3^{(i)}) \in \mathbb{R}^{1 \times 32}, \quad i \in \{1,2,3,4\}
\end{equation}

\subsubsection{Windowed Self-Attention (HAT-1)}

The spatial tokens within each local window $i$, denoted $\mathbf{X}_i \in \mathbb{R}^{4\times32}$ (the flattened patch tokens of that window), are first injected with a learned 2D absolute position embedding $\mathbf{E}_{\text{pos}} \in \mathbb{R}^{4\times32}$:
\begin{equation}
\mathbf{X}_i' = \mathbf{X}_i + \mathbf{E}_{\text{pos}}
\end{equation}

These spatial location-aware tokens and their corresponding carrier token $\mathbf{c}_i$ are concatenated to form the complete token sequence for the window:
\begin{equation}
\mathbf{Z}_i = [\mathbf{X}_i';\, \mathbf{c}_i] \in \mathbb{R}^{5\times32}
\end{equation}

We compute multi-head self-attention using a pre-Layer Normalization (LN) \cite{xiong2020layer} structure with $h=2$ heads and a per-head dimension of $d_h=16$. The queries, keys, and values are derived via learnable linear projections $\mathbf{W}^Q, \mathbf{W}^K, \mathbf{W}^V \in \mathbb{R}^{32 \times 32}$ acting on the layer-normalized tokens $\tilde{\mathbf{Z}}_i = \text{LN}(\mathbf{Z}_i)$:
\begin{equation}
\mathbf{Q}_i, \mathbf{K}_i, \mathbf{V}_i = \tilde{\mathbf{Z}}_i \mathbf{W}^Q,\; \tilde{\mathbf{Z}}_i \mathbf{W}^K,\; \tilde{\mathbf{Z}}_i \mathbf{W}^V
\end{equation}

To provide the attention mechanism with explicit relational awareness, a relative positional bias $\mathbf{B} \in \mathbb{R}^{5 \times 5}$ is added to the attention logits prior to normalization, yielding the attention matrix $\mathbf{A}_i$:
\begin{equation}
\mathbf{A}_i = \text{Softmax}\!\left(\frac{\mathbf{Q}_i \mathbf{K}_i^T}{\sqrt{d_h}} + \mathbf{B} \right)
\end{equation}

The context representations, $\mathbf{A}_i \mathbf{V}_i$ are processed by an output projection $\mathbf{W}^O \in \mathbb{R}^{32 \times 32}$ and integrated via a residual connection. Finally, a feed-forward network (FFN) with an expansion ratio of $r{=}2$ transforms the features:
\begin{align}
\mathbf{Z}_i' &= \mathbf{Z}_i + (\mathbf{A}_i \mathbf{V}_i)\mathbf{W}^O \\
\mathbf{Z}_i'' &= \mathbf{Z}_i' + \text{GELU}\!\left(\text{LN}(\mathbf{Z}_i')\mathbf{P}_1\right)\mathbf{P}_2
\end{align}
where $\mathbf{P}_1 \in \mathbb{R}^{32 \times 64}$ and $\mathbf{P}_2 \in \mathbb{R}^{64 \times 32}$ are the learnable feed-forward weight matrices.

The operations in this block, as illustrated in Fig. \ref{fig:hat_block_detail}, ensure that each spatial token within a local window attends to all other spatial tokens in the same window as well as to its associated carrier token. The carrier token simultaneously aggregates a condensed summary of its window's local spatial details, which is subsequently shared across \emph{all} windows in the carrier token interaction layer.

\subsubsection{Carrier Token Interaction Layer}

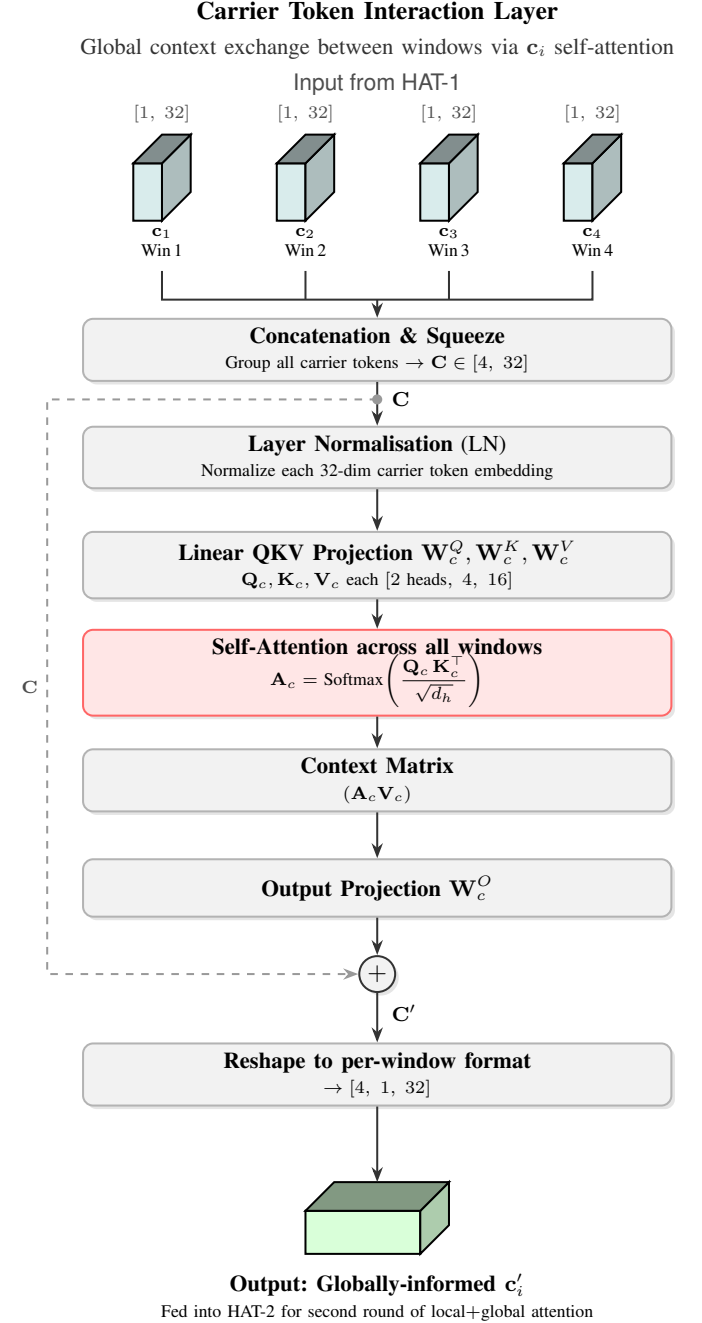
\begin{figure}[!h]
\vspace{-1.5em}
\centering
\resizebox{\columnwidth}{!}{%
\begin{tikzpicture}[
  node distance=0.55cm and 0.35cm,
  process/.style={rectangle, draw=gray!60, fill=gray!10, thick, rounded corners=4pt, minimum height=0.8cm, align=center, font=\small, drop shadow={shadow xshift=0.5mm, shadow yshift=-0.5mm, opacity=0.2}, minimum width=8.2cm},
  attn/.style={rectangle, draw=red!60, fill=red!10, thick, rounded corners=4pt, minimum height=0.8cm, align=center, font=\small, drop shadow={shadow xshift=0.5mm, shadow yshift=-0.5mm, opacity=0.2}, minimum width=8.2cm},
  arrow/.style={-{Stealth[length=6pt]}, thick, draw=black!80},
  skip_arrow/.style={-{Stealth[length=6pt]}, thick, dashed, draw=gray!80},
  lbl/.style={font=\scriptsize\sffamily, text=black!70, align=center},
  add_node/.style={circle, draw=black!80, thick, fill=gray!10, minimum size=0.5cm, inner sep=0pt, font=\normalsize\bfseries, drop shadow={shadow xshift=0.5mm, shadow yshift=-0.5mm, opacity=0.2}}
]
\tikzset{
  pics/cuboid/.style args={w=#1, h=#2, d=#3, fill=#4}{
    code={
      % --- FIX: avoid chained color mixing ---
      \colorlet{basecolor}{#4}
      \colorlet{rightcolor}{basecolor!80!black}
      \colorlet{topcolor}{basecolor!60!black}

      \begin{scope}[shift={(-#1/2 - #3/2, -#2/2 - #3/2)}]
        \draw[fill=basecolor, thick] 
          (0,0) -- ++(#1,0) -- ++(0,#2) -- ++(-#1,0) -- cycle;

        \draw[fill=rightcolor, thick] 
          (#1,0) -- ++(#3,#3) -- ++(0,#2) -- ++(-#3,-#3) -- cycle;

        \draw[fill=topcolor, thick] 
          (0,#2) -- ++(#3,#3) -- ++(#1,0) -- ++(-#3,-#3) -- cycle;

        \coordinate (-west)   at (0, #2/2 + #3/2);
        \coordinate (-east)   at (#1+#3, #2/2 + #3/2);
        \coordinate (-top)    at (#1/2+#3/2, #2+#3);
        \coordinate (-bottom) at (#1/2+#3/2, 0);
        \coordinate (-center) at (#1/2+#3/2, #2/2+#3/2);
      \end{scope}
    }
  }
}
\node[font=\normalsize\bfseries] (title) at (0, 2.3) {Carrier Token Interaction Layer};
\node[font=\small, text=black!80] (sub) at (0, 1.8) {Global context exchange between windows via $\mathbf{c}_i$ self-attention};

\pic (ct1) at (-3, 0) {cuboid={w=0.4, h=0.8, d=0.4, fill=teal!15}};
\node[align=center, font=\scriptsize] at (-3, -0.9) {\textbf{$\mathbf{c}_1$}\\Win\,1};
\node[lbl] at (-3, 0.9) {$[1,\; 32]$};

\pic (ct2) at (-1, 0) {cuboid={w=0.4, h=0.8, d=0.4, fill=teal!15}};
\node[align=center, font=\scriptsize] at (-1, -0.9) {\textbf{$\mathbf{c}_2$}\\Win\,2};
\node[lbl] at (-1, 0.9) {$[1,\; 32]$};

\pic (ct3) at (1, 0) {cuboid={w=0.4, h=0.8, d=0.4, fill=teal!15}};
\node[align=center, font=\scriptsize] at (1, -0.9) {\textbf{$\mathbf{c}_3$}\\Win\,3};
\node[lbl] at (1, 0.9) {$[1,\; 32]$};

\pic (ct4) at (3, 0) {cuboid={w=0.4, h=0.8, d=0.4, fill=teal!15}};
\node[align=center, font=\scriptsize] at (3, -0.9) {\textbf{$\mathbf{c}_4$}\\Win\,4};
\node[lbl] at (3, 0.9) {$[1,\; 32]$};

\node[lbl, font=\small\sffamily] at (0, 1.3) {Input from HAT-1};

\node[process] (reshape) at (0, -2.4) {\textbf{Concatenation \& Squeeze}\\{\scriptsize Group all carrier tokens $\to \mathbf{C} \in [4,\;32]$}};

\node[process] (ln) at (0, -3.9) {\textbf{Layer Normalisation} (LN)\\{\scriptsize Normalize each 32-dim carrier token embedding}};
\node[process] (qkv) at (0, -5.4) {\textbf{Linear QKV Projection} $\mathbf{W}_c^Q, \mathbf{W}_c^K, \mathbf{W}_c^V$\\{\scriptsize $\mathbf{Q}_c,\mathbf{K}_c,\mathbf{V}_c$ each $[2\text{ heads},\; 4,\; 16]$}};
\node[attn] (attn) at (0, -6.9) {\textbf{Self-Attention across all windows}\\{\scriptsize $\mathbf{A}_c = \text{Softmax}\!\left(\dfrac{\mathbf{Q}_c\,\mathbf{K}_c^\top}{\sqrt{d_h}}\right)$ }};
\node[process] (context) at (0, -8.4) {\textbf{Context Matrix}\\{\scriptsize $(\mathbf{A}_c\mathbf{V}_c)$}};
\node[process] (proj) at (0, -9.9) {\textbf{Output Projection} $\mathbf{W}_c^O$};

\node[add_node] (add1) at (0, -11.1) {$+$};

\node[process] (reshape_back) at (0, -12.5) {\textbf{Reshape to per-window format}\\{\scriptsize $\to [4,\;1,\;32]$}};

\pic (out) at (0, -14.5) {cuboid={w=1.6, h=0.6, d=0.4, fill=green!15}};
\node[align=center, font=\small] at (0, -15.6) {\textbf{Output: Globally-informed $\mathbf{c}_i'$}\\{\scriptsize Fed into HAT-2 for second round of local$+$global attention}};

\draw[thick, draw=black!80] (-3, -1.3) -- (-3, -1.7) -- (3, -1.7) -- (3, -1.3);
\draw[thick, draw=black!80] (-1, -1.3) -- (-1, -1.7);
\draw[thick, draw=black!80] (1, -1.3) -- (1, -1.7);

\draw[arrow] (0, -1.7) -- (reshape.north);

\draw[arrow] (reshape.south) -- coordinate[pos=0.4] (res_start) node[right=2pt, pos=0.4, font=\footnotesize\sffamily] {$\mathbf{C}$} (ln.north);
\draw[arrow] (ln.south) -- (qkv.north);
\draw[arrow] (qkv.south) -- (attn.north);
\draw[arrow] (attn.south) -- (context.north);
\draw[arrow] (context.south) -- (proj.north);
\draw[arrow] (proj.south) -- (add1.north);

\draw[arrow] (add1.south) -- node[right=2pt, pos=0.4, font=\footnotesize\sffamily] {$\mathbf{C}'$} (reshape_back.north);
\draw[arrow] (reshape_back.south) -- (out-top);

\fill[gray!80] (res_start) circle (2pt);
\draw[skip_arrow] (res_start) -- ++(-4.6,0) |- node[lbl, left, pos=0.25] {$\mathbf{C}$} (add1.west);

\end{tikzpicture}%
}
\caption{Internal processing of the carrier token interaction layer. The four carrier tokens $\mathbf{c}_1,\ldots,\mathbf{c}_4$ are grouped as $\mathbf{C} \in \mathbb{R}^{4\times32}$ and exchange global information via a $4{\times}4$ self-attention matrix ($\mathbf{A}_c$) before being projected and distributed back to their respective windows as $\mathbf{c}_i'$.}
\label{fig:ct_interaction_detail}
\vspace{0.3\baselineskip}
\end{figure}

As illustrated in Fig.~\ref{fig:ct_interaction_detail}, the carrier tokens facilitate global context propagation by exchanging information across all spatial windows via self-attention. The carrier tokens are grouped and squeezed into $\mathbf{C} \in \mathbb{R}^{4\times32}$, then layer-normalized to $\tilde{\mathbf{C}} = \text{LN}(\mathbf{C})$.

The interaction between carrier tokens is formulated as an attention operation followed by mapping the tokens back to their embedding space without an additional FFN. The operations use learned projection matrices $\mathbf{W}_c^Q, \mathbf{W}_c^K, \mathbf{W}_c^V, \mathbf{W}_c^O \in \mathbb{R}^{32 \times 32}$ as follows:
\begin{align}
\mathbf{Q}_c, \mathbf{K}_c, \mathbf{V}_c &= \tilde{\mathbf{C}}\mathbf{W}_c^Q,\; \tilde{\mathbf{C}}\mathbf{W}_c^K,\; \tilde{\mathbf{C}}\mathbf{W}_c^V \\
\mathbf{A}_c &= \text{Softmax}\!\left(\frac{\mathbf{Q}_c \mathbf{K}_c^T}{\sqrt{d_h}}\right) \\
\mathbf{C}' &= \mathbf{C} + (\mathbf{A}_c\mathbf{V}_c)\mathbf{W}_c^O
\end{align}

After the carrier token interaction layer, a second identical HAT block (\textit{HAT-2}) computes attention between the globally-enriched carrier tokens $\mathbf{c}_i'$ and the spatial features.

\subsection{Classification Head with QP Conditioning}

Following the dual HAT block processing, the refined spatial tokens are rearranged into a 2D feature map to aggregate the learned context features. This final spatial representation, which excludes the auxiliary carrier tokens used for context exchange, is denoted as:
\begin{equation}
\mathbf{F}_{\text{out}} \in \mathbb{R}^{4 \times 4 \times 32}.
\end{equation}

To condense the spatial information into a global descriptor, Global Average Pooling (GAP) is applied across the spatial dimensions. This yields a 32-dimensional feature vector:
\begin{equation}
\mathbf{f}_s = \text{GAP}(\mathbf{F}_{\text{out}}) \in \mathbb{R}^{32}.
\end{equation}

To ensure the model makes partition decisions that are sensitive to the target bitrate and quality, the quantization parameter (QP) is incorporated into the prediction logic. The QP value $q$ is normalized to the range $[0, 1]$ and concatenated with the spatial features to form a fused feature vector:
\begin{equation}
\tilde{\mathbf{f}} = [\mathbf{f}_s;\, q/51] \in \mathbb{R}^{33}.
\end{equation}

The classification is performed by a three-layer multilayer perceptron (MLP) head equipped with BN \cite{ioffe2015batch} and dropout \cite{srivastava2014dropout} regularization to improve generalization. The first fully connected (FC) layer expands the fused features to $d_1$ neurons with a dropout rate of $p_1$. The second FC layer further projects the features to $d_2$ neurons with a dropout rate of $p_2$. Finally, the third FC layer has 21 output neurons with logistic activations representing the split likelihoods for every potential CU across the three depth levels:
\begin{align}
\mathbf{h}_1 &= \text{Dropout}\!\left(\text{ReLU}\!\left(\text{BN}(\mathbf{W}_1^T\,\tilde{\mathbf{f}})\right),\; p_1\right) \\
\mathbf{h}_2 &= \text{Dropout}\!\left(\text{ReLU}\!\left(\text{BN}(\mathbf{W}_2^T\,\mathbf{h}_1)\right),\; p_2\right) \\
\hat{\mathbf{y}} &= \text{Sigmoid}(\mathbf{W}_3^T\,\mathbf{h}_2) \in [0,1]^{21}
\end{align}
where $\mathbf{W}_1 \in \mathbb{R}^{d_1 \times 33}$, $\mathbf{W}_2 \in \mathbb{R}^{d_2 \times d_1}$, and $\mathbf{W}_3 \in \mathbb{R}^{21 \times d_2}$ are learnable weight matrices. The dimensions $d_1$ and $d_2$ are varied to control the number of trainable parameters of the model as per the number of training samples available at different resolutions to achieve a good fit. The dropout rates $p_1$ and $p_2$ decrease progressively, reflecting the reduced overfitting risk as the representation compresses toward the 21-dimensional output.

This QP-conditioned head allows the architecture to jointly reason about spatial texture complexity and the target compression level, enabling a single trained model to predict CU partitions across the entire QP range.

\subsection{Fused Batch Normalization}

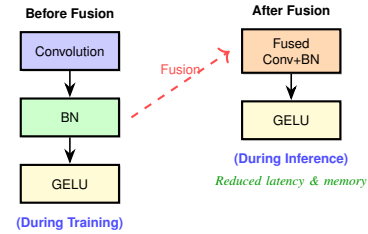
\begin{figure}[htbp]
\centering
\begin{tikzpicture}[
    node distance=0.35cm,
    module/.style={
        rectangle, draw=black, thick,
        minimum width=1.3cm, minimum height=0.5cm,
        align=center, font=\tiny\sffamily,
        drop shadow={shadow xshift=0.2mm, shadow yshift=-0.2mm, opacity=0.25}
    },
    arrow/.style={->, >=Stealth, semithick},
    phase/.style={font=\tiny\sffamily\bfseries, text=blue!70}
]
\node[module, fill=blue!20] (conv1) {Convolution};
\node[module, fill=green!20, below=of conv1] (bn1) {BN};
\node[module, fill=yellow!20, below=of bn1] (act1) {GELU};
\draw[arrow] (conv1) -- (bn1);
\draw[arrow] (bn1) -- (act1);
\node[above=0.05cm of conv1, font=\tiny\sffamily\bfseries] {Before Fusion};
\node[phase, below=0.05cm of act1] {(During Training)};
\node[module, fill=orange!30, right=1.6cm of conv1] (fused) {Fused\\Conv+BN};
\node[module, fill=yellow!20, below=of fused] (act2) {GELU};
\draw[arrow] (fused) -- (act2);
\node[above=0.05cm of fused, font=\tiny\sffamily\bfseries] {After Fusion};
\node[phase, below=0.05cm of act2] {(During Inference)};
\draw[->, thick, red!70, dashed] ([xshift=0.15cm]bn1.east) -- ([xshift=-0.15cm]fused.west)
    node[midway, above, font=\tiny\sffamily] {Fusion};
\node[below=0.35cm of act2, font=\tiny\itshape, text=green!60!black] {Reduced latency \& memory};
\end{tikzpicture}
\caption{Layer fusion process: batch normalization parameters $(\gamma, \beta, \mu, \sigma^2)$ are absorbed into convolutional weights $\mathbf{W}$ post-training, producing fused weights $\hat{\mathbf{W}}$ and $\hat{b}$ via BN fusion, eliminating redundant operations during inference.}
\label{fig:fusion}
\end{figure}

During inference, we fuse BN into preceding layers following structural reparameterization conventions. Figure~\ref{fig:fusion} illustrates how the separate convolution--BN--activation chain used during training is collapsed into a single fused convolution--activation unit at inference time, eliminating the redundant BN forward pass and reducing memory bandwidth.

For a convolution with weight $\mathbf{W}$ and bias $b$, and BN parameters $(\gamma,\, \beta,\, \mu,\, \sigma^2)$, where $\gamma$ is the learnable per-channel scale, $\beta$ is the learnable per-channel shift, $\mu$ is the running mean accumulated during training, and $\sigma^2$ is the corresponding running variance, the fused parameters are computed as:
\begin{align}
\hat{\mathbf{W}} &= \frac{\gamma}{\sqrt{\sigma^2 + \epsilon}}\, \mathbf{W} \\
\hat{b} &= \frac{\gamma (b - \mu)}{\sqrt{\sigma^2 + \epsilon}} + \beta
\end{align}
where $\epsilon$ is a small positive constant of the order of $10^{-6}$. 

Similar fusion is also performed for the linear FC layers in the classification head, directly improving overall inference latency.

\section{Experimental Results}
\label{sec:experiments}

\subsection{Experimental Setup}

Table~\ref{tab:experimental_setup} summarizes the experimental configuration, including the dataset, encoder version, training hyperparameters, and the training and inference hardwares used for all experiments reported in this work.

\begin{table}[htbp]
\centering
\caption{Experimental Configuration}
\label{tab:experimental_setup}
\begin{tabular}{ll}
\toprule
\textbf{Component} & \textbf{Specification} \\
\midrule
Dataset Name & CPH-Intra~\cite{li2018cph} \\
Encoder & HM 16.5, intra-mode \\
\midrule
\multicolumn{2}{l}{\textit{Training Hardware}} \\
GPU & NVIDIA A40 (46 GB VRAM) \\
CPU & 2$\times$ Intel Xeon Silver 4509Y (32 Cores / 64 Threads) \\
\midrule
\multicolumn{2}{l}{\textit{Training Hyperparameters}} \\
Optimizer & AdamW \\
Learning Rate & $1\times10^{-4}$ \\
Weight Decay & 0.01 \\
Batch Size & 64 \\
Scheduler & Cosine Annealing \\
Max Epochs & 10,000 \\
Loss Function & Hierarchical Binary Cross-Entropy \\
\midrule
\multicolumn{2}{l}{\textit{Inference System}} \\
CPU & AMD Ryzen 5 7235HS (3.2 GHz, 4C/8T) \\
RAM & 12 GB DDR4 \\
\bottomrule
\end{tabular}
\end{table}

The hierarchical binary cross-entropy loss employed for training 
reflects the conditional structure of the HEVC quad-tree. At the 
$64{\times}64$ level (L$_1$), every CTU always has a valid split 
decision, so a standard binary cross-entropy is applied. At the 
$32{\times}32$ (L$_2$) and $16{\times}16$ (L$_3$) levels, a binary 
validity mask derived from the parent block's ground-truth split 
decision ensures that only semantically meaningful partition decisions 
contribute to the loss. Concretely, if a $64{\times}64$ CTU is not 
split at L$_1$, all four corresponding $32{\times}32$ decisions at 
L$_2$ and all sixteen $16{\times}16$ decisions at L$_3$ are masked out 
and contribute nothing to the training objective. The total loss is the 
equally weighted sum of the three level-wise losses, reflecting the 
design choice that partition errors at any depth level incur equivalent 
rate-distortion cost.

To ensure representative evaluation, experiments were conducted across four resolution subsets of the CPH-Intra dataset~\cite{li2018cph}, \cite{li2017deep} which contains CTUs and their respective ground truth partitions derived from the brute-force RDO of the HM 16.5 encoder at the four reference QPs $\{22, 27, 32, 37\}$ and at four different resolutions spanning from 768$\times$512 to 4928$\times$3264. The dataset for each resolution is derived by downsampling 2000 images (partitioned into 1700 images for training, 100 for validation and 200 for testing) from a resolution of 4928$\times$3264 to the target resolution~\cite{xu2018reducing}. All three models (HFViT, ETH-CNN, and H-FCN) were trained separately for each of the four resolutions (4928$\times$3264, 2880$\times$1920, 1536$\times$1024, and 768$\times$512), with a single unified model trained across four QPs $\{22, 27, 32, 37\}$ at each resolution. Table~\ref{tab:dataset_splits} details the number of CTU samples allocated to training, validation, and testing for each resolution.

\begin{table}[htbp]
\centering
\caption{CPH-Intra Dataset Split Sizes (Number of CTU Samples) Across Resolutions}
\label{tab:dataset_splits}
\resizebox{\columnwidth}{!}{%
\begin{tabular}{l | r | r | r | r}
\toprule
\textbf{Resolution} & \textbf{Train} & \textbf{Valid} & \textbf{Test} & \textbf{Total} \\
\midrule
4928$\times$3264 & 1,668,975 & 98,175 & 196,350 & 1,963,500 \\
2880$\times$1920 & 573,750 & 33,750 & 67,500 & 675,000 \\
1536$\times$1024 & 163,200 & 9,600 & 19,200 & 192,000 \\
768$\times$512 & 40,800 & 2,400 & 4,800 & 48,000 \\
\bottomrule
\end{tabular}%
}
\end{table}

\subsection{Partition Prediction Accuracy}
\label{subsec:accuracy}

Table~\ref{tab:accuracy_results} presents the partition prediction accuracy of all three models across four resolutions evaluated on the test partition of the CPH-Intra dataset. Accuracy is reported at the levels of $64 \times 64$ (L\textsubscript{1}), $32 \times 32$ (L\textsubscript{2}), $16 \times 16$ (L\textsubscript{3}) CUs. The overall accuracy across the three levels of CU partitions is also reported.

We compared the performance of HFViT with ETH-CNN \cite{xu2018reducing} and H-FCN \cite{paul2020speeding}, which are CNN based architectures introduced for the task of accelerating intra-mode partition prediction. Specifically, the ETH-CNN \cite{xu2018reducing} model achieved benchmark results on the CPH-Intra dataset, while the H-FCN \cite{paul2020speeding} introduced an extremely lightweight bottom-up hierarchical fully convolutional architecture that achieved competitive performance. Table \ref{tab:accuracy_results} also reports the number of trainable parameters for the different models at each resolution. Since the number of CTUs available in the CPH-Intra dataset varies across different resolutions, we reduced the number of parameters of ETH-CNN \cite{xu2018reducing} and HFViT, keeping the depth and structure of the models fixed to reduce overfitting whenever necessary while training at the lower resolutions. For ETH-CNN \cite{xu2018reducing}, the parameter reduction at lower resolutions was achieved by reducing the number of output channels in each convolutional layer uniformly across all three branches, which in turn reduced the dimensionality of the concatenated feature vector fed to the fully connected layers, with the hidden layer widths of all three fully connected branches scaled down proportionally. For HFViT, the parameter reduction was achieved solely by reducing the hidden dimensions $d_1$ and $d_2$ of the two intermediate fully connected layers in the QP-conditioned classification head, while all other architectural components, including the convolutional backbone, HAT blocks, and carrier token interaction layer, were kept unchanged. The H-FCN \cite{paul2020speeding} has substantially fewer parameters, so its architecture was kept fixed across resolutions.

In terms of overall accuracy across the three levels of partition, HFViT consistently outperformed ETH-CNN~\cite{xu2018reducing} at every resolution and depth level, while H-FCN had the best accuracy at level L\textsubscript{3} of the partition tree. However, H-FCN struggled with predicting accurate partition decisions at higher resolutions as evinced by its wider performance gaps with the other models at 4928$\times$3264 and 2880$\times$1920. This suggests that the limited learning capacity and bottom-up architecture of the H-FCN model is not effective at capturing the fine grained texture variations that are crucial for partition decision at higher resolutions. HFViT achieved the best overall accuracy across all three partition levels and resolutions.

\begin{table}[htbp]
\centering
\caption{Partition Prediction Accuracy (\%) Across Models, Resolutions, and Depth Levels}
\label{tab:accuracy_results}
{\footnotesize
\setlength{\tabcolsep}{3pt}
\begin{tabular}{l|l|r|cccc}
\toprule
\textbf{Model} & \textbf{Resolution} & \textbf{Parameters} & \textbf{L\textsubscript{1}} & \textbf{L\textsubscript{2}} & \textbf{L\textsubscript{3}} & \textbf{Overall} \\
\midrule
\multirow{4}{*}{ETH-CNN~\cite{xu2018reducing}}
  & 4928$\times$3264 & 1,288,210 & 85.90 & 80.42 & 77.61 & 81.31 \\
  & 2880$\times$1920 & 1,288,210 & 82.32 & 80.56 & 75.96 & 79.61 \\
  & 1536$\times$1024 & 1,288,210 & 88.84 & 81.16 & 73.55 & 81.18 \\
  & 768$\times$512   &   401,494 & 88.73 & 75.53 & 65.27 & 76.51 \\
\midrule
\multirow{4}{*}{H-FCN~\cite{paul2020speeding}}
  & 4928$\times$3264 &    19,187 & 72.87 & 76.24 & \textbf{80.67} & 76.59 \\
  & 2880$\times$1920 &    19,187 & 79.51 & 78.12 & \textbf{80.20} & 79.27 \\
  & 1536$\times$1024 &    19,187 & 84.75 & 83.90 & \textbf{81.91} & 83.52 \\
  & 768$\times$512   &    19,187 & 90.25 & 85.87 & \textbf{82.87} & 86.33 \\
\midrule
\multirow{4}{*}{\textbf{HFViT}}
  & 4928$\times$3264 & 1,670,794 & \textbf{86.36} & \textbf{81.56} & 79.51 & \textbf{82.48} \\
  & 2880$\times$1920 & 1,670,794 & \textbf{88.73} & \textbf{82.90} & 78.34 & \textbf{83.32} \\
  & 1536$\times$1024 & 1,318,666 & \textbf{93.51} & \textbf{86.37} & 79.13 & \textbf{86.34} \\
  & 768$\times$512   &   454,282 & \textbf{94.94} & \textbf{87.12} & 79.70 & \textbf{87.25} \\
\bottomrule
\end{tabular}
}
\end{table}

\subsection{Rate-Distortion Performance}
To evaluate RD performance, we integrated the trained HFViT model
with the reference HM encoder such that the RDO based CU partition decisions in the intra
mode were replaced by the predictions of HFViT. Subsequently, the HM encoder integrated
with the HFViT model was used to perform intra-mode encoding of several test sequences.
The quality of the decoded frames was measured using PSNR, MS-SSIM \cite{wang2003multiscale} and VMAF \cite{li2016vmaf}, where the
latter two are widely used to evaluate perceptual quality in the video coding industry. The RD performance of the videos encoded using CU partitions predicted by HFViT was quantified by computing the Bj{\o}ntegaard delta rate (BD-rate) \cite{bjontegaard2001calculation} metric with the encodes
obtained by using the unmodified HM encoder as reference. The encoding speedup over the HM Encoder
is defined as:
\begin{equation}
\text{Speedup} = \left(1 - \frac{T_{\text{Model}}}{T_{\text{ref}}}\right) \times 100\%,
\label{eq:speedup}
\end{equation}
where $T_{\text{ref}}$ is the encoding time of the reference unmodified HM encoder and
$T_{\text{Model}}$ is the encoding time of the HM encoder integrated with the
proposed model. The BD-rate and speedup computations for ETH-CNN and H-FCN also follow the same procedure.

\subsubsection{Performance on Test Set}

\begin{figure}[htb]
\centering
\includegraphics[width=0.9\columnwidth]{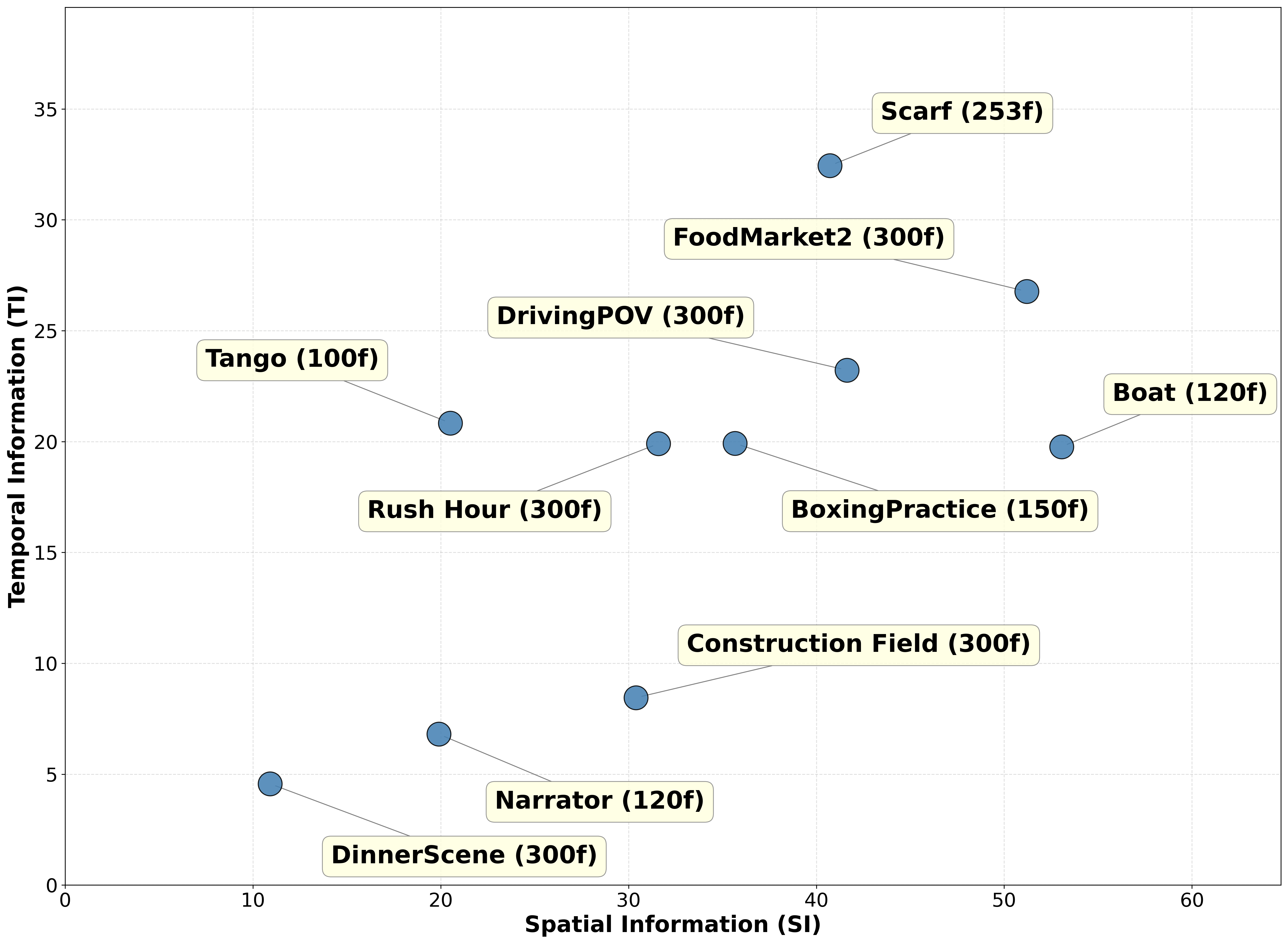}
\caption{Spatial Information (SI) vs.\ Temporal Information (TI)~\cite{itu1999subjective} of the ten test sequences used in the experiments. The sequences are at $3840\times2160$ and $4096\times2160$ resolution, 4:2:0 chroma subsampling, and 8-bit depth. Each point is annotated with the sequence name and its total frame count.}
\label{fig:si_ti_plot}
\end{figure}

To evaluate the performance of the HFViT model, we compiled a test set consisting of ten uncompressed sequences spanning a wide range of spatial information (SI) and temporal information (TI)~\cite{itu1999subjective,robitza2021impact} values, as shown in Fig.~\ref{fig:si_ti_plot}, to ensure adequate content diversity. The wide spread across both axes in Fig.~\ref{fig:si_ti_plot} confirms that the test set covers low-motion static scenes as well as high-motion dynamic content with varying texture complexity, ensuring that the reported results are not biased toward any particular content type. These ten sequences, which are widely used for video coding performance evaluation, were originally at $3840\times2160$ and $4096\times2160$ resolutions in YUV 4:4:4 format at 10-bit depth, which were subsequently converted to YUV 4:2:0 format at 8-bit depth, and were downsampled to $2880\times1920$, $1536\times1024$, and $768\times512$ for the evaluation at the respective lower resolutions. For evaluation at the resolution of $4928 \times 3264$, the frames of the test sequences were padded by replicating the edge pixels. 

% Since all three evaluation resolutions are exact multiples of the $64\times64$ CTU tile size, each frame is partitioned directly into a uniform grid of $64\times64$ tiles without any border padding. Should a test sequence at an arbitrary resolution not satisfy this divisibility condition, the inference pipeline would ceil-pad the frame dimensions to the nearest multiple of 64, filling the border regions via edge-pixel replication to avoid introducing artificial discontinuities, and subsequently discard the predictions corresponding to the padded tiles. 

Table~\ref{tab:confidence_intervals} reports the average BD-rate and speedup obtained per resolution. To quantify the reliability of our experimental results, the corresponding 95\% confidence intervals (CIs) are also reported in Table~\ref{tab:confidence_intervals}. The narrow confidence intervals for HFViT across all metrics confirm that its performance gains are statistically significant and consistent across diverse content types.

Table~\ref{tab:confidence_intervals} reveals that the proposed HFViT model attained the best RD performance with respect to the perceptual VMAF quality metric across all four resolutions. Its RD performances in terms of PSNR and MS-SSIM were also better than both ETH-CNN and H-FCN at all resolutions except at $768 \times 512$, where H-FCN achieved the lowest average BD-rates for these two metrics. However, at the highest resolution, H-FCN struggled to predict partitions accurately, as already pointed out in Section \ref{subsec:accuracy}, consequently impairing its RD performance. Considering the tradeoff between BD-rate and speedup, our model achieved the most impressive gains at the highest resolution of 4928$\times$3264, which is quite attractive considering the current preponderance of video content at UHD or higher resolutions. At lower resolutions, HFViT consistently achieved superior RD performance while attaining competitive speedups.

\begin{table*}[!t]
\centering
\caption{BD-rate and Speed-up with 95\% Confidence Intervals (CI) Metrics Across Resolutions and Models (10 test sequences per resolution).}
\label{tab:confidence_intervals}
{\footnotesize
\setlength{\tabcolsep}{4pt}
\renewcommand{\arraystretch}{1.15}
\begin{tabular}{l l r@{\,}c@{\,}l r@{\,}c@{\,}l r@{\,}c@{\,}l r@{\,}c@{\,}l}
\toprule
\multirow{2}{*}{\textbf{Resolution}}
  & \multirow{2}{*}{\textbf{Model}}
  & \multicolumn{3}{c}{\textbf{BD-rate PSNR (\%)}}
  & \multicolumn{3}{c}{\textbf{BD-rate VMAF (\%)}}
  & \multicolumn{3}{c}{\textbf{BD-rate MS-SSIM (\%)}}
  & \multicolumn{3}{c}{\textbf{Speedup (\%)}} \\
\cmidrule(lr){3-5}\cmidrule(lr){6-8}\cmidrule(lr){9-11}\cmidrule(lr){12-14}
  & & Mean & & [95\% CI] & Mean & & [95\% CI] & Mean & & [95\% CI] & Mean & & [95\% CI] \\
\midrule
\multirow{3}{*}{4928$\times$3264}
  & ETH-CNN~\cite{xu2018reducing}
    & 4.59 & & {[3.65,\;5.53]}
    & 4.96 & & {[3.87,\;6.05]}
    & 5.07 & & {[3.89,\;6.25]}
    & 78.57 & & {[76.71,\;80.44]} \\
  & H-FCN~\cite{paul2020speeding}
    & 10.18 & & {[3.27,\;17.09]}
    & 10.59 & & {[3.97,\;17.21]}
    & 10.72 & & {[4.07,\;17.37]}
    & 59.96 & & {[57.46,\;62.47]} \\
  & \textbf{HFViT}
    & \textbf{2.86} & & \textbf{[2.33,\;3.39]}
    & \textbf{2.81} & & \textbf{[2.15,\;3.47]}
    & \textbf{3.25} & & \textbf{[2.24,\;4.27]}
    & \textbf{79.41} & & \textbf{[77.92,\;80.90]} \\
\midrule
\multirow{3}{*}{2880$\times$1920}
  & ETH-CNN~\cite{xu2018reducing}
    & 6.07 & & {[4.10,\;8.04]}
    & 5.96 & & {[3.95,\;7.98]}
    & 6.81 & & {[3.81,\;9.81]}
    & \textbf{76.19} & & \textbf{[69.74,\;82.64]} \\
  & H-FCN~\cite{paul2020speeding}
    & 4.54 & & {[2.60,\;6.47]}
    & 4.47 & & {[2.53,\;6.41]}
    & 5.19 & & {[2.25,\;8.12]}
    & 66.34 & & {[61.28,\;71.41]} \\
  & \textbf{HFViT}
    & \textbf{3.27} & & \textbf{[2.20,\;4.33]}
    & \textbf{3.07} & & \textbf{[2.23,\;3.91]}
    & \textbf{3.66} & & \textbf{[1.51,\;5.81]}
    & 75.73 & & {[70.55,\;80.91]} \\
\midrule
\multirow{3}{*}{1536$\times$1024}
  & ETH-CNN~\cite{xu2018reducing}
    & 7.73 & & {[5.39,\;10.07]}
    & 7.68 & & {[5.07,\;10.30]}
    & 8.44 & & {[6.00,\;10.88]}
    & \textbf{75.14} & & \textbf{[69.62,\;80.65]} \\
  & H-FCN~\cite{paul2020speeding}
    & 4.47 & & {[2.48,\;6.47]}
    & 4.43 & & {[2.59,\;6.28]}
    & 4.94 & & {[3.10,\;6.79]}
    & 54.60 & & {[51.09,\;58.11]} \\
  & \textbf{HFViT}
    & \textbf{4.07} & & \textbf{[2.74,\;5.40]}
    & \textbf{3.94} & & \textbf{[2.74,\;5.13]}
    & \textbf{4.41} & & \textbf{[3.19,\;5.62]}
    & 64.27 & & {[59.08,\;69.46]} \\
\midrule
\multirow{3}{*}{768$\times$512}
  & ETH-CNN~\cite{xu2018reducing}
    & 9.22 & & {[4.45,\;13.98]}
    & 7.95 & & {[3.99,\;11.91]}
    & 9.04 & & {[4.11,\;13.98]}
    & \textbf{66.67} & & \textbf{[56.21,\;77.13]} \\
  & H-FCN~\cite{paul2020speeding}
    & \textbf{2.89} & & \textbf{[1.90,\;3.89]}
    & 3.55 & & {[2.51,\;4.59]}
    & \textbf{3.84} & & \textbf{[2.69,\;4.99]}
    & 40.83 & & {[31.53,\;50.13]} \\
  & \textbf{HFViT}
    & 3.35 & & {[2.27,\;4.44]}
    & \textbf{3.46} & & \textbf{[2.43,\;4.49]}
    & 4.15 & & {[3.16,\;5.14]}
    & 46.56 & & {[40.38,\;52.75]} \\
\bottomrule
\end{tabular}
}
\end{table*}

\subsubsection{Performance on JCT-VC Test Sequences}

We also evaluated the proposed model on the JCT-VC test set~\cite{bossen2010common} which is the standardized set of test sequences widely used to benchmark HEVC codec performance. Table~\ref{tab:all_results} reports BD-rate (for PSNR, VMAF, MS-SSIM~\cite{wang2003multiscale}) and encoding speedup across the different test classes. The key observations from this table are enumerated as follows:

\begin{enumerate}
    \item HFViT achieved the best overall rate-distortion trade-off, with average VMAF BD-rates of 3.12\%, 3.36\%, and 6.03\% for Classes A, B, and E respectively. The BD-rates attained by HFViT for both the perceptual metrics (MS-SSIM and VMAF) consistently outperformed ETH-CNN~\cite{xu2018reducing} and H-FCN across all three classes. HFViT also achieved a lower PSNR BD-rate than the two CNN based models for Class B and Class E sequences, while for Class A sequences H-FCN gave the lowest PSNR BD-rates.
    \item HFViT delivered superior speedup for Class A sequences and competitive speedups for Class B and Class E sequences as compared to the ETH-CNN model. Across all three classes, H-FCN, despite its lightweight architecture, attained lower speedup than the proposed HFViT model.
    \item For Class E sequences, all models incurred higher BD-rate penalties than at the other classes, which can be attributed to the limited availability of similar content in the training dataset. Yet, HFViT significantly reduced the average BD-rates across all metrics as compared to ETH-CNN. While H-FCN~\cite{paul2020speeding} also achieved comparable BD-rates, HFViT offers superior speedups than H-FCN for Class E sequences.
\end{enumerate}

\begin{table*}[!t]
\centering
\caption{Performance Comparison of ETH-CNN, H-FCN and proposed HFViT on JCT-VC Test sequences}
\label{tab:all_results}
\renewcommand{\arraystretch}{1.3}
\resizebox{\textwidth}{!}{%
\large
\begin{tabular}{@{}l l rrrr rrrr rrrr@{}}
\toprule
\multirow{2}{*}{\textbf{Class}} &
\multirow{2}{*}{\textbf{Test Sequence}} &
\multicolumn{4}{c}{\textbf{ETH-CNN~\cite{xu2018reducing}}} &
\multicolumn{4}{c}{\textbf{H-FCN~\cite{paul2020speeding}}} &
\multicolumn{4}{c}{\textbf{HFViT}} \\
\cmidrule(lr){3-6} \cmidrule(lr){7-10} \cmidrule(lr){11-14}
& &
\textbf{BD-rate} & \textbf{BD-rate} & \textbf{BD-rate} & \textbf{Speed-up} &
\textbf{BD-rate} & \textbf{BD-rate} & \textbf{BD-rate} & \textbf{Speed-up} &
\textbf{BD-rate} & \textbf{BD-rate} & \textbf{BD-rate} & \textbf{Speed-up} \\
& &
\textbf{PSNR (\%)} & \textbf{VMAF (\%)} & \textbf{MS-SSIM (\%)} & \textbf{(\%)} &
\textbf{PSNR (\%)} & \textbf{VMAF (\%)} & \textbf{MS-SSIM (\%)} & \textbf{(\%)} &
\textbf{PSNR (\%)} & \textbf{VMAF (\%)} & \textbf{MS-SSIM (\%)} & \textbf{(\%)} \\
\midrule
\multirow{5}{*}{\shortstack[l]{Class A\\(2560$\times$1600)}}
& SteamLocomotiveTrain & 4.72 & 6.26 & 5.51 & 80.45 & 3.06 & 3.58 & 3.55 & 76.29 & 2.54 & 2.25 & 3.00 & 80.33 \\
& Traffic              & 0.42 & 3.42 & 1.66 & 40.91 & 1.24 & 1.08 & 0.78 & 40.15 & 2.28 & 0.32 & 0.07 & 73.03 \\
& PeopleOnStreet       & 0.50 & 1.01 & 1.40 & 37.61 & 1.42 & 1.83 & 1.56 & 40.29 & 2.62 & 1.46 & 0.60 & 67.56 \\
& NebutaFestival       & 3.65 & 11.40 & 3.87 & 76.38 & 1.86 & 8.84 & 3.13 & 76.07 & 1.76 & 8.43 & 2.40 & 78.18 \\
\cmidrule(lr){2-14}
& \textbf{Average}     & \textbf{2.32} & \textbf{5.52} & \textbf{3.11} & \textbf{58.83} &
                         \textbf{1.89} & \textbf{3.83} & \textbf{2.25} & \textbf{58.20} &
                         \textbf{2.30} & \textbf{3.12} & \textbf{1.52} & \textbf{74.77} \\
\midrule
\multirow{6}{*}{\shortstack[l]{Class B\\(1920$\times$1080)}}
& BasketballDrive    & 8.74 & 8.80 & 9.32 & 75.01 & 5.01 & 5.46 & 5.61 & 58.31 & 4.55 & 4.71 & 4.84 & 66.17 \\
& BQTerrace          & 3.77 & 5.25 & 5.68 & 64.51 & 2.30 & 3.51 & 3.39 & 50.81 & 2.27 & 3.19 & 3.28 & 55.65 \\
& Cactus             & 6.06 & 6.36 & 7.52 & 73.32 & 3.25 & 3.40 & 4.19 & 52.17 & 3.24 & 3.56 & 4.21 & 58.67 \\
& Kimono1            & 4.75 & 4.60 & 4.91 & 81.77 & 4.18 & 3.44 & 3.91 & 63.71 & 3.15 & 2.47 & 3.06 & 74.46 \\
& ParkScene          & 4.59 & 4.82 & 5.50 & 74.57 & 2.35 & 2.44 & 3.95 & 53.67 & 2.77 & 2.86 & 4.28 & 62.78 \\
\cmidrule(lr){2-14}
& \textbf{Average}   & \textbf{5.58} & \textbf{5.96} & \textbf{6.59} & \textbf{73.83} &
                       \textbf{3.42} & \textbf{3.65} & \textbf{4.21} & \textbf{55.73} &
                       \textbf{3.19} & \textbf{3.36} & \textbf{3.94} & \textbf{63.55} \\
\midrule
\multirow{4}{*}{\shortstack[l]{Class E\\(1280$\times$720)}}
& Johnny             & 18.42 & 17.74 & 19.17 & 79.43 & 7.76 & 7.42 & 8.49 & 58.45 & 8.81 & 7.60 & 9.46 & 67.20 \\
& KristenAndSara     & 14.43 & 13.41 & 15.47 & 75.29 & 7.13 & 6.72 & 7.85 & 54.63 & 6.12 & 5.50 & 6.18 & 66.22 \\
& FourPeople         & 12.14 & 10.62 & 13.58 & 66.02 & 5.50 & 5.43 & 7.76 & 49.12 & 5.09 & 4.98 & 6.04 & 60.06 \\
\cmidrule(lr){2-14}
& \textbf{Average}   & \textbf{15.00} & \textbf{13.93} & \textbf{16.08} & \textbf{73.58} &
                       \textbf{6.80}  & \textbf{6.52}  & \textbf{8.03}  & \textbf{54.07} &
                       \textbf{6.67}  & \textbf{6.03}  & \textbf{7.23}  & \textbf{64.49} \\
\bottomrule
\end{tabular}%
}
\end{table*}

\subsubsection{Rate-Distortion Curves}

\begin{figure*}[!t]
\centering
\begin{minipage}[b]{0.33\textwidth}
    \centering
    \includegraphics[width=\textwidth]{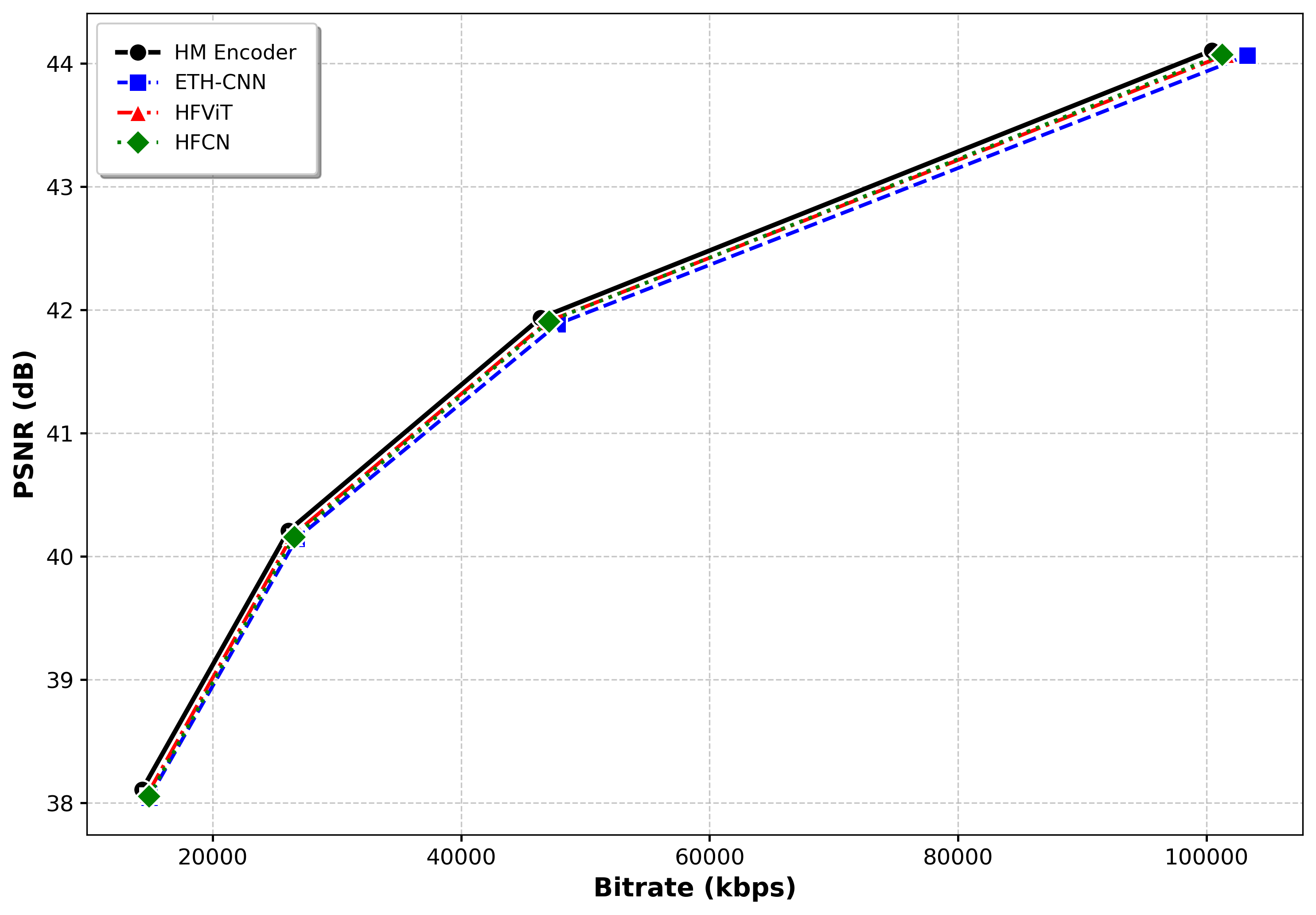}
\end{minipage}%
\hspace{0.001\textwidth}%
\begin{minipage}[b]{0.33\textwidth}
    \centering
    \includegraphics[width=\textwidth]{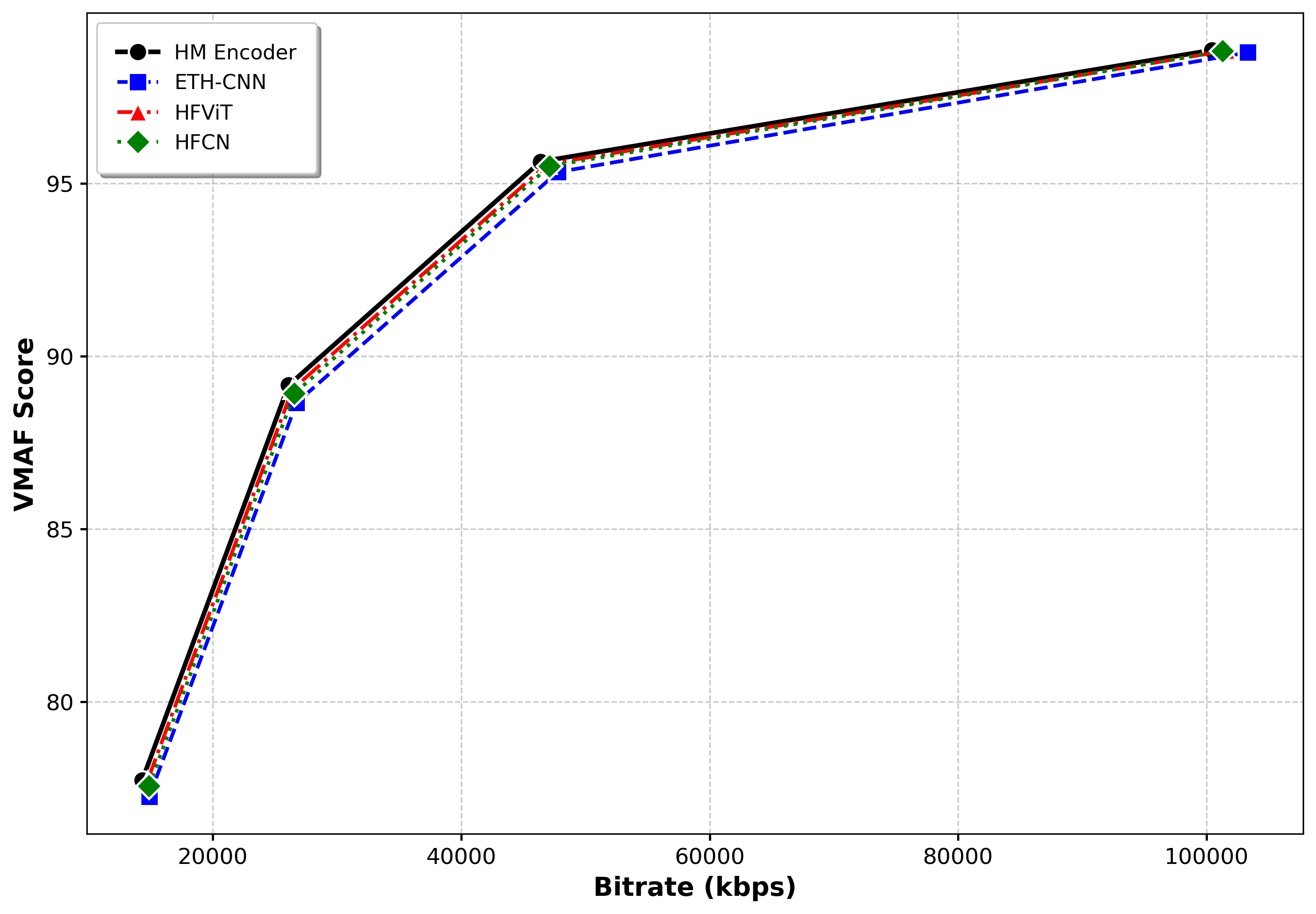}
\end{minipage}%
\hspace{0.001\textwidth}%
\begin{minipage}[b]{0.33\textwidth}
    \centering
    \includegraphics[width=\textwidth]{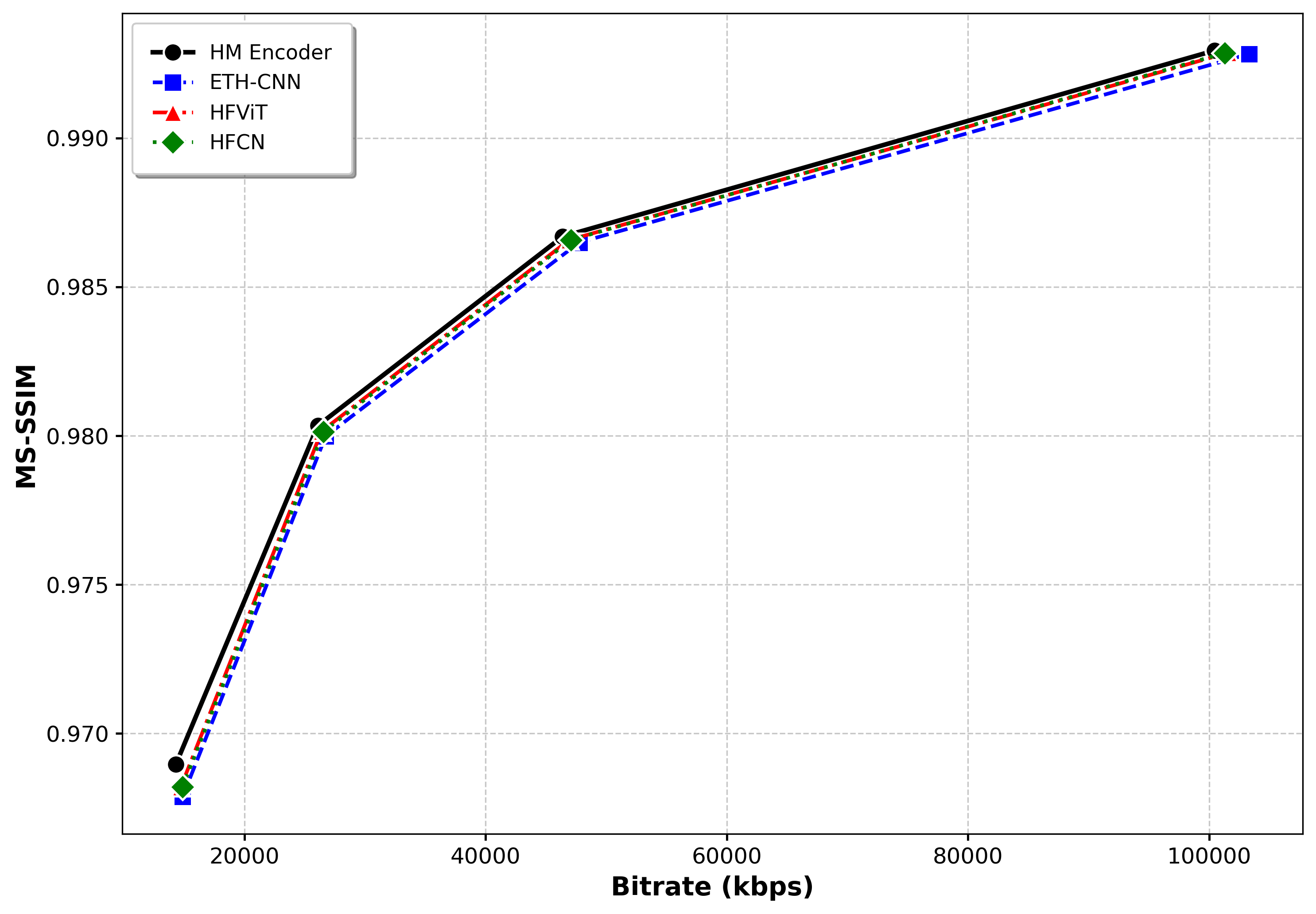}
\end{minipage}
\vspace{0.4em}
\begin{minipage}[b]{0.33\textwidth}
    \centering
    \includegraphics[width=\textwidth]{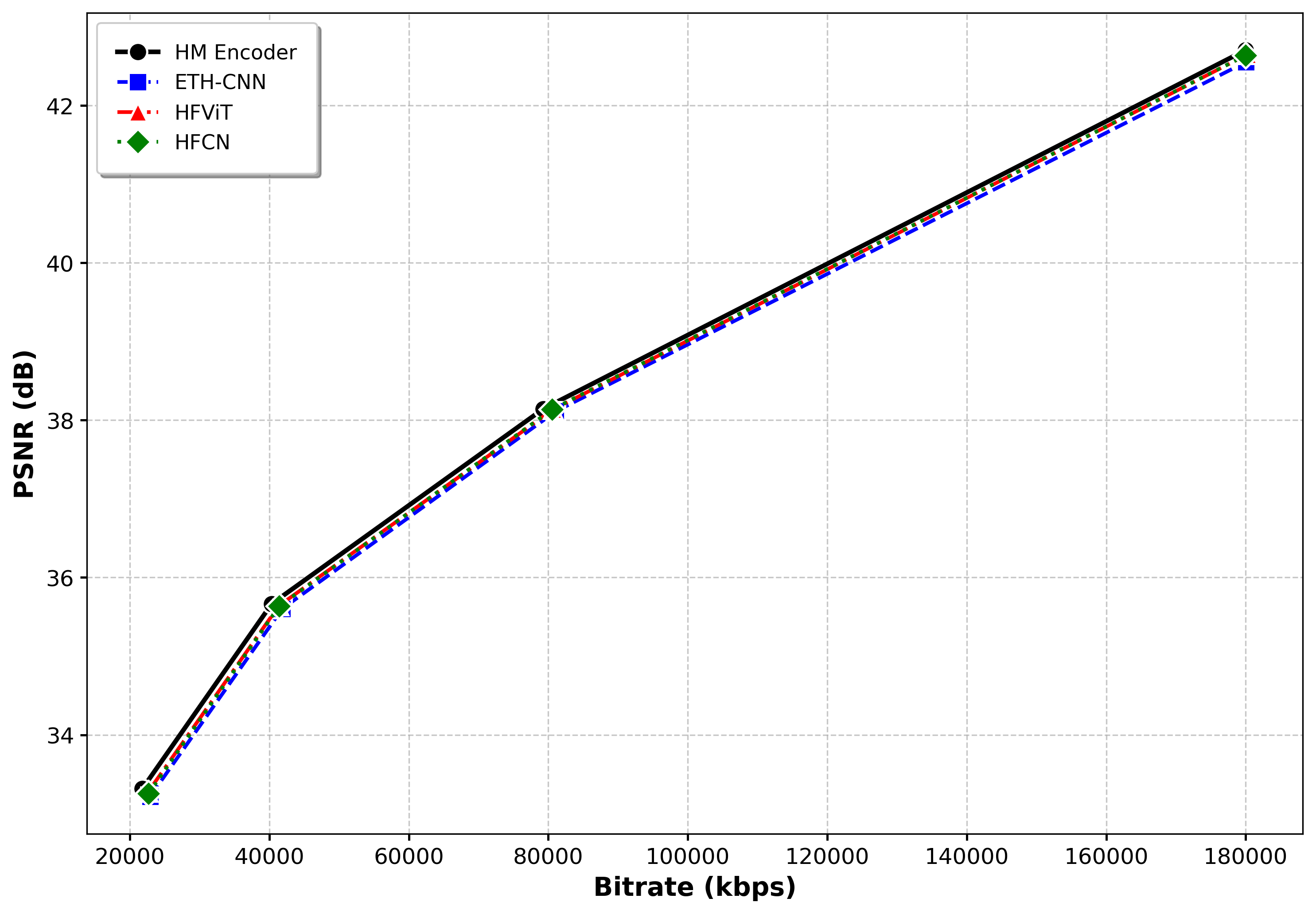}
\end{minipage}%
\hspace{0.001\textwidth}%
\begin{minipage}[b]{0.33\textwidth}
    \centering
    \includegraphics[width=\textwidth]{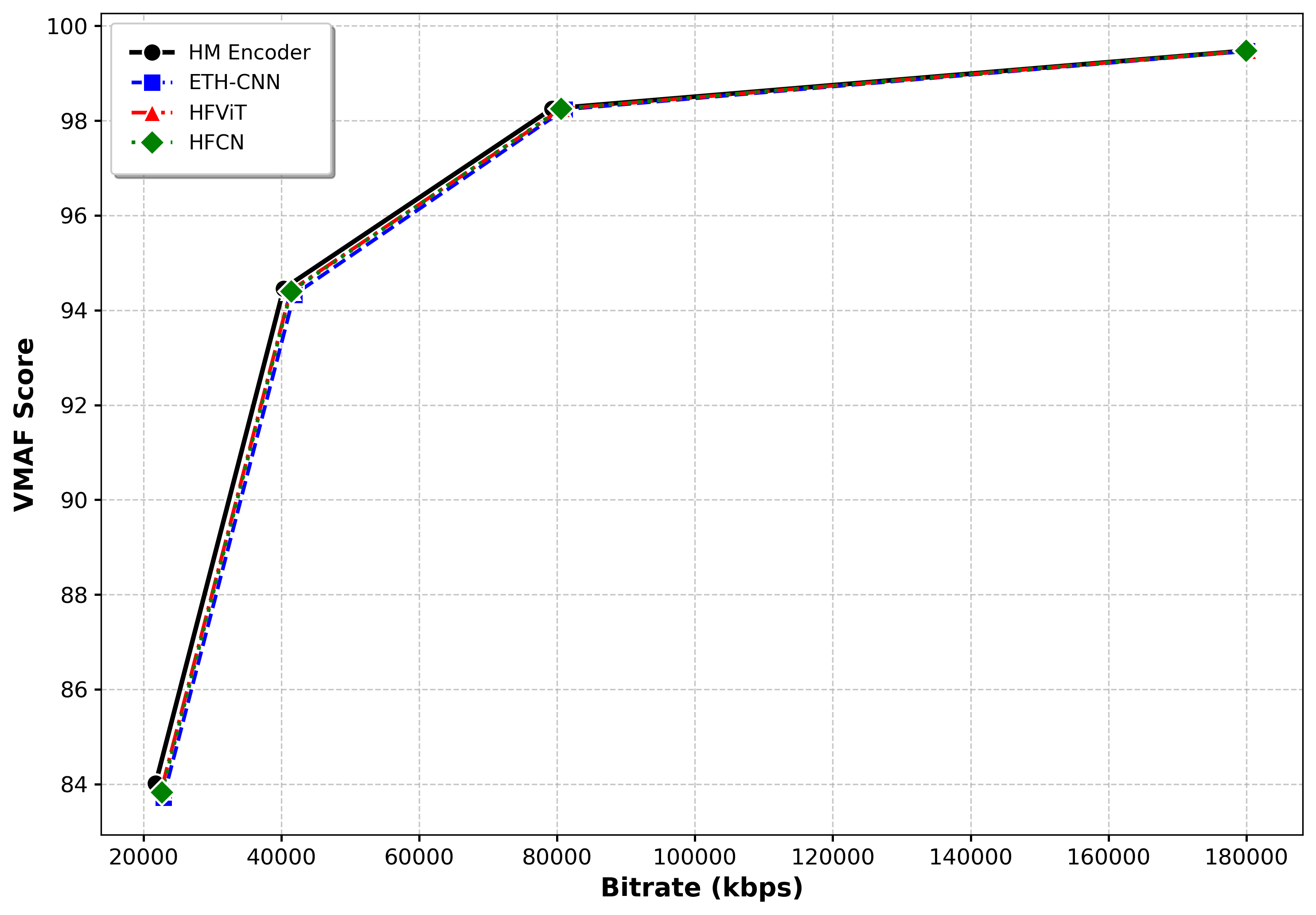}
\end{minipage}%
\hspace{0.001\textwidth}%
\begin{minipage}[b]{0.33\textwidth}
    \centering
    \includegraphics[width=\textwidth]{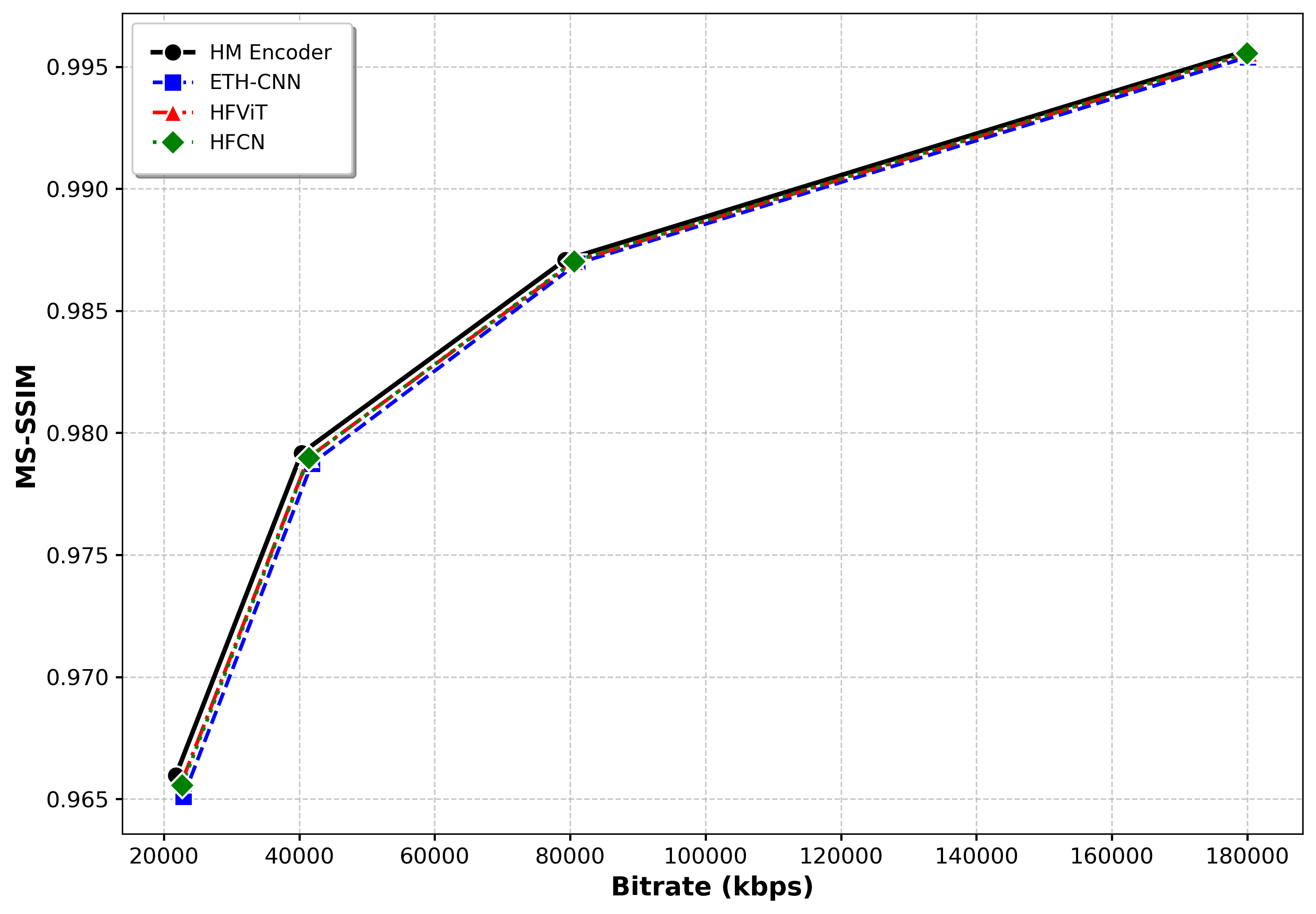}
\end{minipage}
\vspace{0.4em}
\begin{minipage}[b]{0.33\textwidth}
    \centering
    \includegraphics[width=\textwidth]{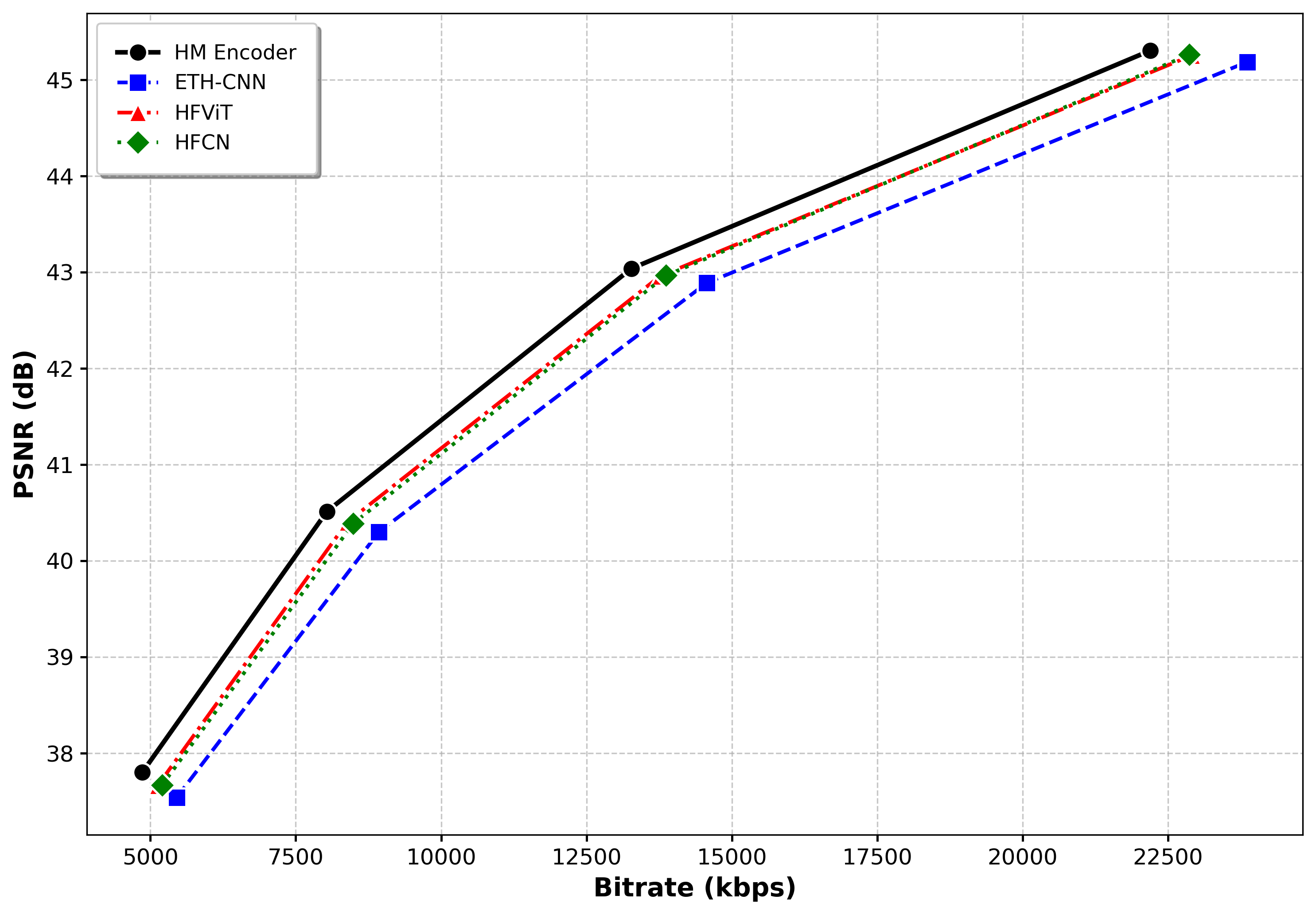}
\end{minipage}%
\hspace{0.001\textwidth}%
\begin{minipage}[b]{0.33\textwidth}
    \centering
    \includegraphics[width=\textwidth]{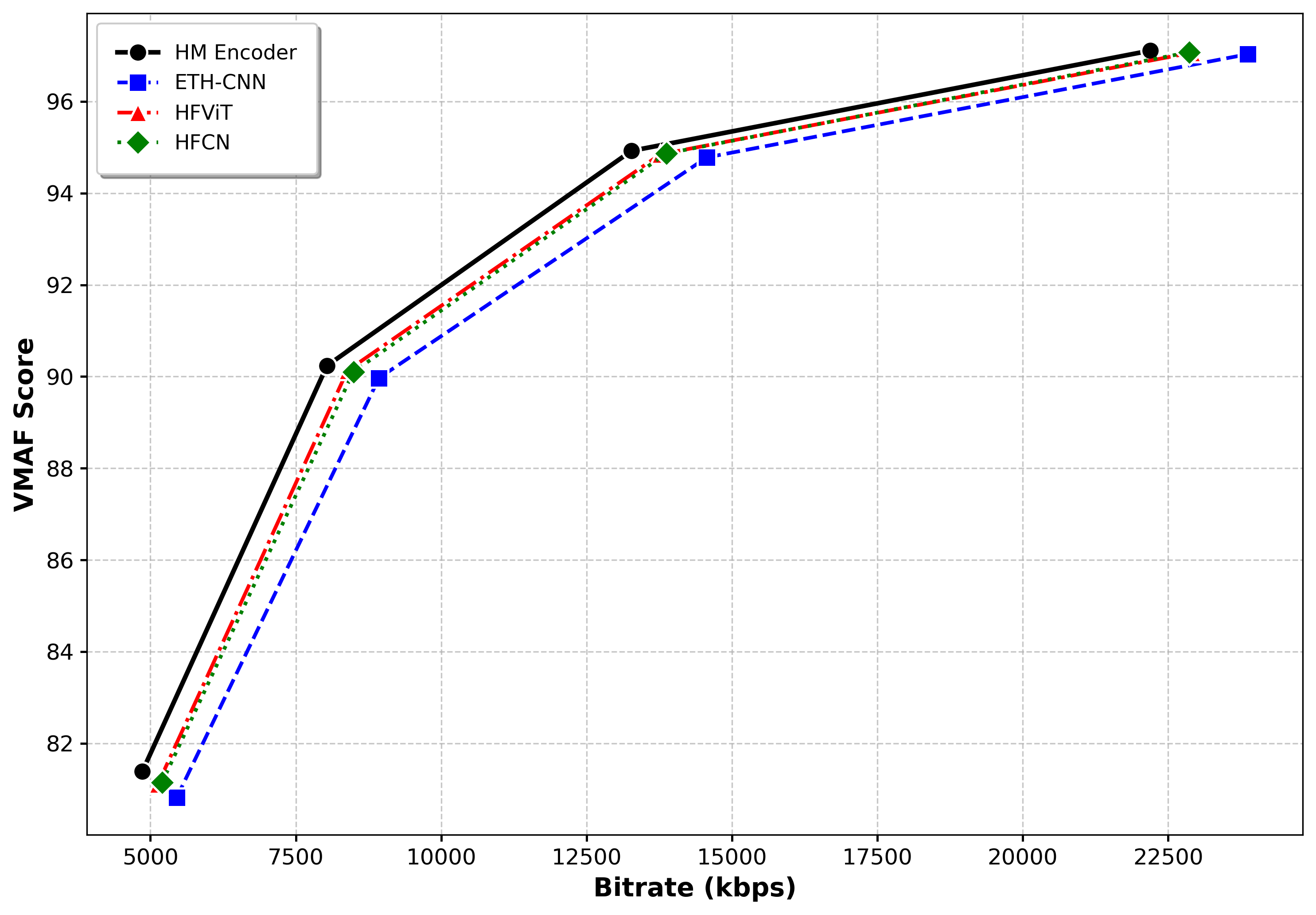}
\end{minipage}%
\hspace{0.001\textwidth}%
\begin{minipage}[b]{0.33\textwidth}
    \centering
    \includegraphics[width=\textwidth]{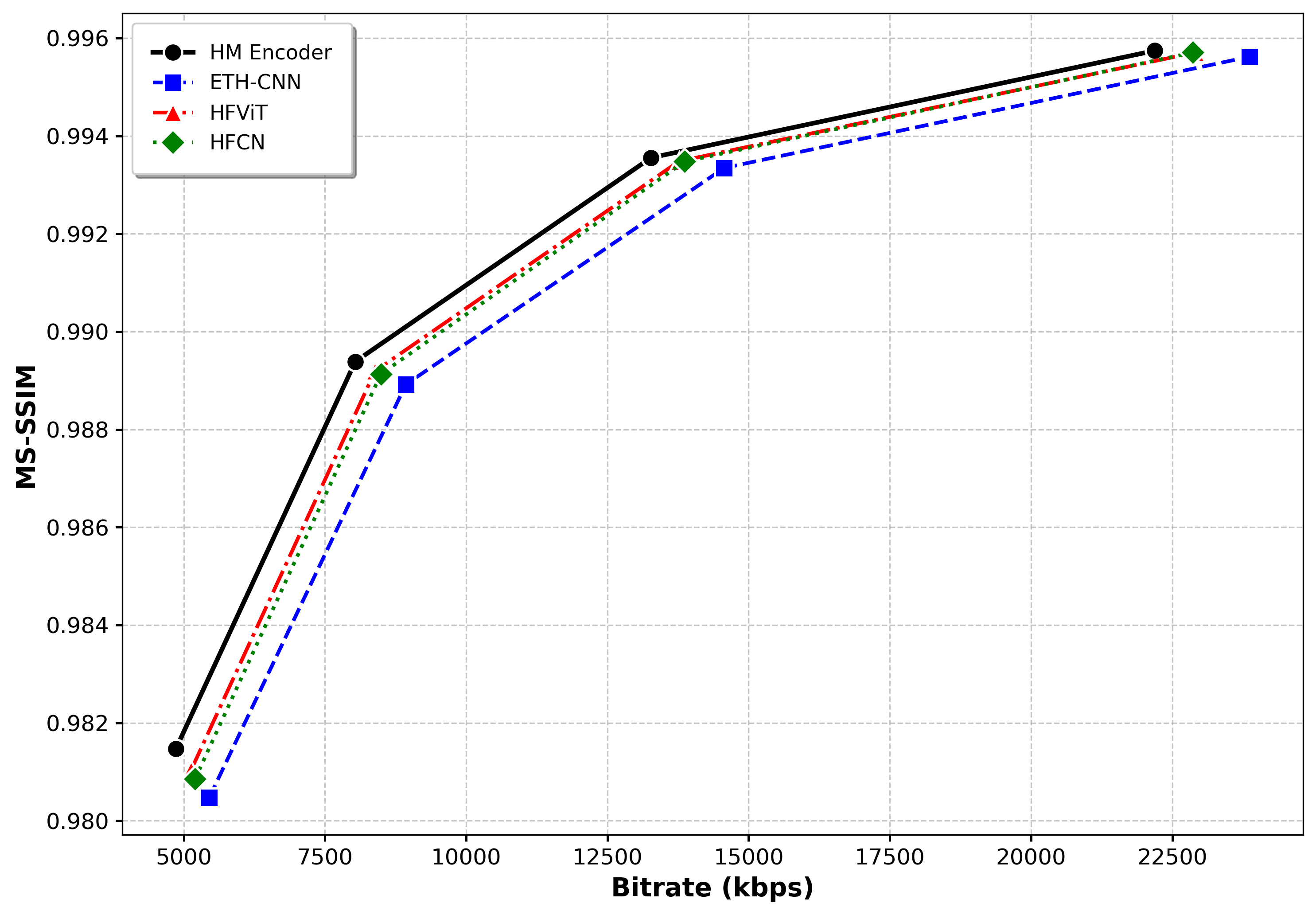}
\end{minipage}
\caption{%
    Rate-distortion curves comparing HFViT,
    ETH-CNN~\cite{xu2018reducing}, and H-FCN~\cite{paul2020speeding}
    against the HM~16.5 anchor across three video
    sequences and three quality metrics.
    \textbf{Row~1:} \textit{SteamLocomotiveTrain}
    (Class~A, 2560$\times$1600).
    \textbf{Row~2:} \textit{BQTerrace}
    (Class~B, 1920$\times$1080).
    \textbf{Row~3:} \textit{KristenAndSara}
    (Class~E, 1280$\times$720).
    Columns present PSNR (left), VMAF (centre),
    and MS-SSIM (right) plotted against bitrate (kbps)
    at QP $\in \{22, 27, 32, 37\}$.
    HFViT consistently tracks closest to the HM~16.5
    anchor across all sequences and metrics,
    outperforming both ETH-CNN~\cite{xu2018reducing} and
    H-FCN~\cite{paul2020speeding}.%
}
\label{fig:rd_curves_all}
\end{figure*}

To visualize the compression efficiency trade-offs, Fig.~\ref{fig:rd_curves_all} presents comprehensive
RD curves comparing HFViT,
ETH-CNN~\cite{xu2018reducing}, and H-FCN~\cite{paul2020speeding}
against the HM encoder on three representative
test sequences: \textit{SteamLocomotiveTrain} (Class~A,
2560$\times$1600), \textit{BQTerrace} (Class~B,
1920$\times$1080), and \textit{KristenAndSara}
(Class~E, 1280$\times$720). Each row corresponds to
one video sequence, and each column shows a different
quality metric---PSNR, VMAF, and MS-SSIM---plotted
against bitrate (kbps) at QP values $\{22, 27, 32, 37\}$. The RD curves visually corroborate the quantitative findings
from Table~\ref{tab:all_results}, validating that HFViT achieves the closest approximation to the RDO based reference encoder across all evaluated quality metrics and video classes.

% \begin{figure*}[!t]
% \centering
% \begin{subfigure}{0.48\textwidth}
% \centering
% \includegraphics[width=\textwidth]{frame17.drawio.png}
% \caption{QP~37}
% \label{fig:partition_f17}
% \end{subfigure}
% \hfill
% \begin{subfigure}{0.48\textwidth}
% \centering
% \includegraphics[width=\textwidth]{frame2.drawio.png}
% \caption{QP~32}
% \label{fig:partition_f22}
% \end{subfigure}
% \caption{CU partition visualization on \texttt{IntraValid\_1536$\times$1024}: (a) QP~37 and (b) QP~32. Each panel shows from top to bottom: Ground truth from HM~16.5 RDO (\textcolor{red}{red}), HFViT prediction (\textcolor{blue}{blue}), ETH-CNN~\cite{xu2018reducing} prediction (\textcolor{green}{green}), and H-FCN~\cite{paul2020speeding} prediction (\textcolor{magenta}{magenta}). HFViT most accurately reproduces fine-grained depth-2 and depth-3 partition boundaries, particularly in transitional texture regions.}
% \label{fig:partition_comparison}
% \end{figure*}

\begin{figure*}[!t]
\centering
\subfloat[QP~37]{
    \includegraphics[width=0.48\textwidth]{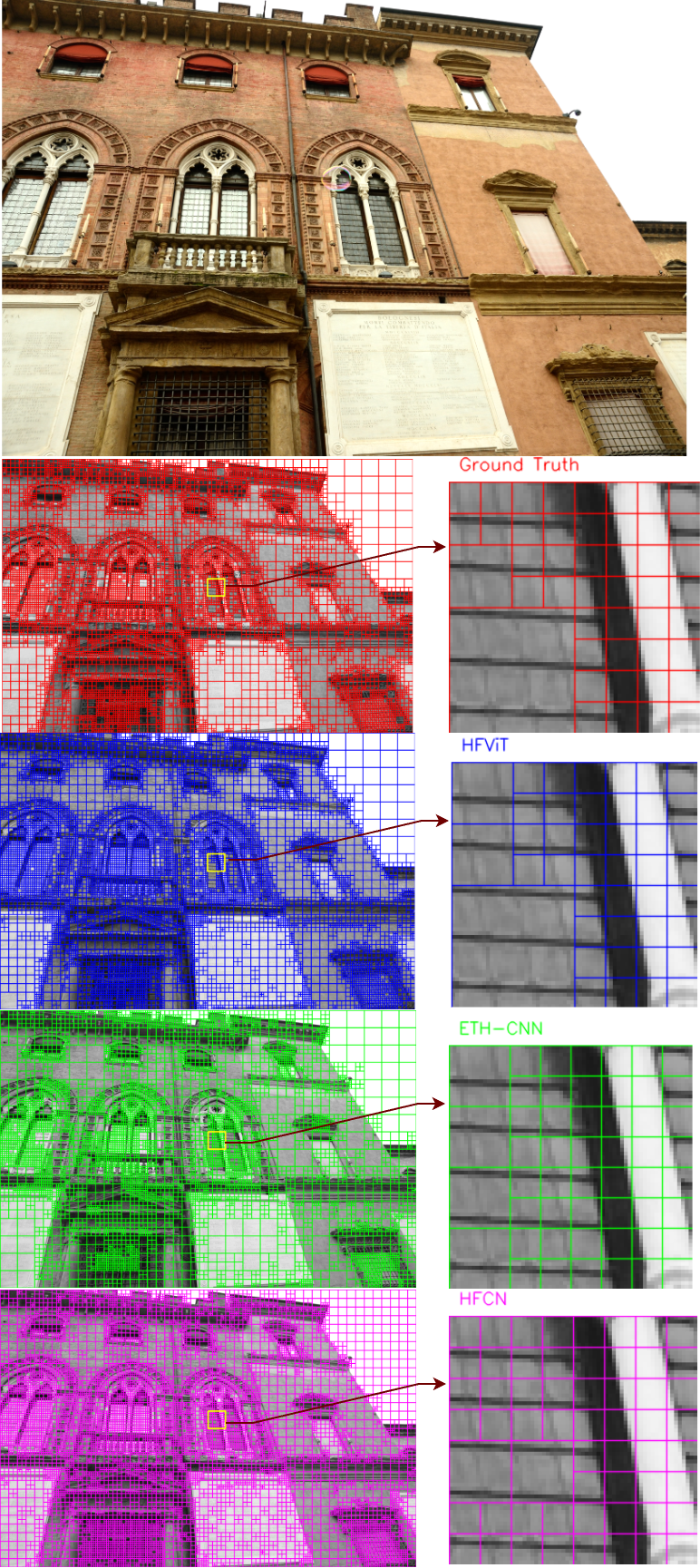}
    \label{fig:partition_f17}
}
\hfill
\subfloat[QP~32]{
    \includegraphics[width=0.48\textwidth]{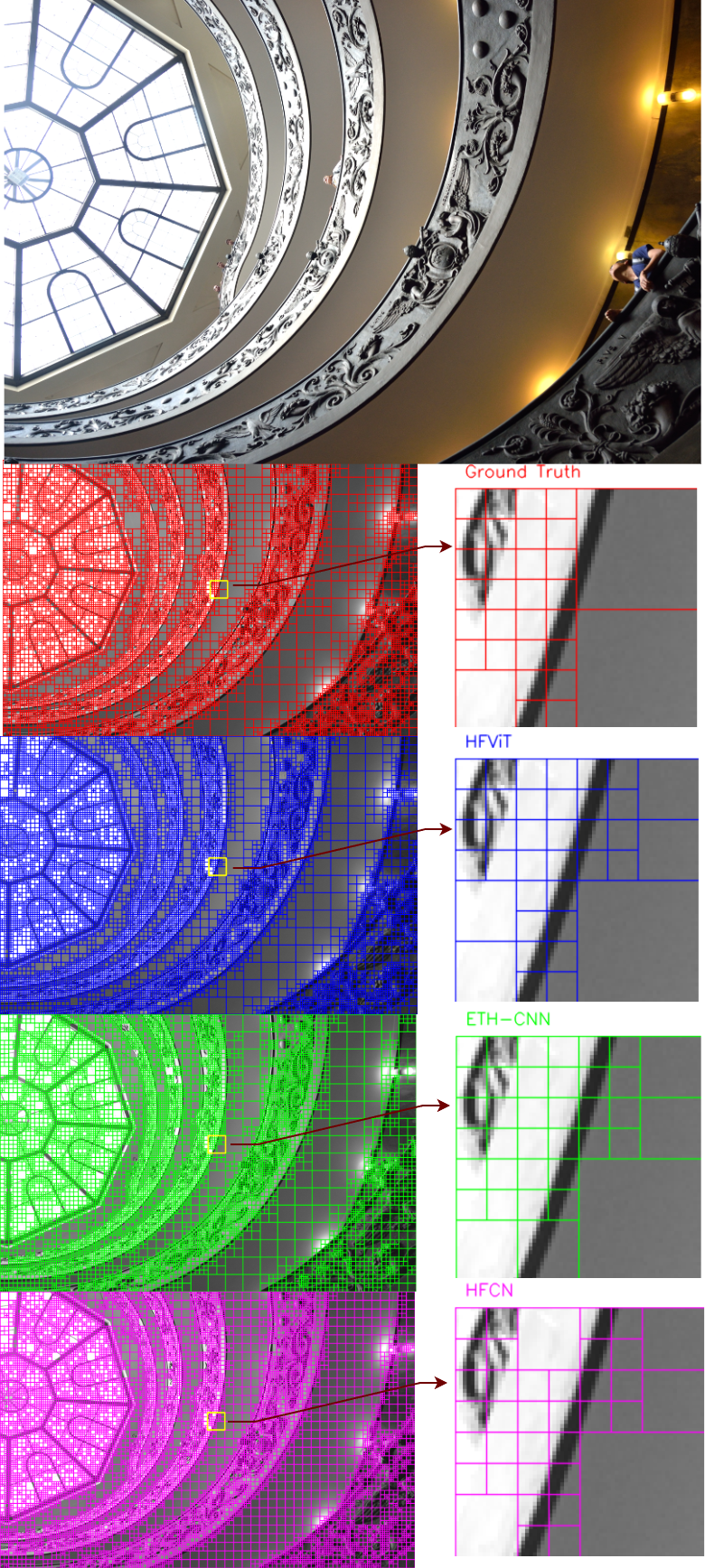}
    \label{fig:partition_f22}
}
\caption{CU partition visualization on \texttt{IntraValid\_1536$\times$1024}: 
(a) QP~37 and (b) QP~32. Each panel shows from top to bottom: Ground truth 
from HM~16.5 RDO (\textcolor{red}{red}), HFViT prediction 
(\textcolor{blue}{blue}), ETH-CNN~\cite{xu2018reducing} prediction 
(\textcolor{green}{green}), and H-FCN~\cite{paul2020speeding} prediction 
(\textcolor{magenta}{magenta}). HFViT most accurately reproduces fine-grained 
depth-2 and depth-3 partition boundaries, particularly in transitional 
texture regions.}
\label{fig:partition_comparison}
\end{figure*}

\subsection{Qualitative Partition Visualization}

While Table~\ref{tab:accuracy_results} summarizes accuracy numerically, it cannot reveal \emph{where} partition errors occur spatially. To complement the quantitative results, Fig. ~\ref{fig:partition_comparison} provides a visual comparison of the predicted CU partition maps against the ground truth derived by the HM encoder's RDO on two representative frames at 1536$\times$1024 taken from the validation partition of the CPH-Intra dataset. These examples are chosen to illustrate model behaviour under different encoding conditions: Frame~17 at QP~37 (high quantization, coarser partitions) and Frame~22 at QP~32 (moderate quantization, finer structure). Visually inspecting the predicted boundaries reveals how well each model reproduces the encoder's depth-level decisions, particularly in regions with complex textures or abrupt transitions, where HFViT's global context modeling provides the most visible advantage over the purely local CNN baselines.

\subsection{Ablation Studies}

To individually assess the contribution of each architectural choice made to design HFViT,
Table~\ref{tab:ablation} presents an ablation study conducted on the
1536$\times$1024 resolution subset, evaluating three configurations:
the full HFViT model, HFViT without BN fusion,
and HFViT without depthwise separable convolutions.

\begin{table}[htbp]
\centering
\caption{Impact of BN Fusion and Depthwise Separable Convolution on Performance (1536$\times$1024 Resolution)}
\label{tab:ablation}
\resizebox{\columnwidth}{!}{%
\begin{tabular}{l c c c c}
\toprule
\textbf{\large Model Variant} &
\textbf{\large \shortstack{BD-rate\\PSNR (\%)}} &
\textbf{\large \shortstack{BD-rate\\VMAF (\%)}} &
\textbf{\large \shortstack{BD-rate\\MS-SSIM (\%)}} &
\textbf{\large \shortstack{Speed-up\\wrt HM (\%)}} \\
\midrule
\large HFViT
  & \large 4.07 & \large 3.94 & \large 4.41 & \large \textbf{64.27} \\
\large HFViT w/o BN Fusion
  & \large 4.07 & \large 3.94 & \large 4.41 & \large 55.56 \\
\large HFViT w/o Depthwise Sep.\ Conv.\
  & \large \textbf{3.33} & \large \textbf{3.25} & \large \textbf{3.90} & \large 55.56 \\
\bottomrule
\end{tabular}
}
\end{table}

Two key observations emerge from Table~\ref{tab:ablation}. First, BN fusion acts as a strictly lossless runtime optimization; it yielded an 8.71\% improvement in inference speedup (from 55.56\% to 64.27\%) without altering partition predictions and hence the BD-rate. Second, replacing depthwise separable convolutions with regular convolutions increased the model size from 1,318,666 to 1,335,265 trainable parameters and improved compression efficiency (reducing BD-rate penalties by 0.5--0.7\%) at the expense of reducing the speedup by 8.71\%. This confirms that utilizing depthwise separable convolutions effectively trades a marginal drop in accuracy for a leaner model and significantly faster CPU inference.

\section{Conclusion}
\label{sec:conclusion}

This paper addressed a key challenge of incorporating long-range spatial dependencies in learning to predict HEVC partitions while meeting the stringent CPU latency requirements of practical encoder pipelines by introducing HFViT, a hybrid architecture that unifies a reparameterized depthwise-separable convolutional backbone with a hierarchical attention mechanism augmented by carrier tokens for sub-quadratic global context propagation. Extensive evaluation on four resolutions and multiple quality metrics demonstrated superior RD performance when compared against competing CNN-based architectures such as ETH-CNN \cite{xu2018reducing} and H-FCN \cite{paul2020speeding}, while maintaining attractive speedups, with the most compelling speedups attained at higher resolutions which are more challenging to encode from a complexity standpoint. 

The inherent ability of attention mechanisms to capture long-range dependencies could be further exploited by implementing cross-frame hierarchical attention to predict inter mode partitions for HEVC, which constitutes the primary focus of our future work. Further, we plan to benchmark the proposed HFViT model on more advanced codecs such as AV1 and VVC, which have far more allowed partition types that compound the challenge of partition decision due to the combinatorial complexity of the search space. The ability of HFViT to model global context is likely to enable significant acceleration of the corresponding encoders by successfully learning complex partition decisions at such high dimensional search spaces. It is also interesting to apply the extended HFViT model with cross-frame attention to learn diverse encoding decisions such as merge candidate selection and motion vector prediction. 

\ifCLASSOPTIONcaptionsoff
  \newpage
\fi
\bibliographystyle{IEEEtran}
\bibliography{references}

\end{document}